\documentclass{aa}
\usepackage{txfonts}
\usepackage{graphicx}
\usepackage{lipsum}
\usepackage{subcaption}        
\usepackage{lscape}             
\usepackage{placeins}
\usepackage[version=4]{mhchem}
\usepackage[colorlinks=true, linkcolor=blue, citecolor=blue, urlcolor=blue]{hyperref}
\usepackage{balance}

\begin{document}
\raggedbottom
   \title{Metal-Poor Stars as Metal-Rich Impostors via Planet Engulfment}

   \author{Zongheng Li\inst{1,3}\email{lizongheng@ynao.ac.cn}
        \and Shi Jia\inst{1,2,3}
        \and Jie Su\inst{1,3}
        \and Shang-Fei Liu\inst{4}
        \and Jianheng Guo\inst{1,3}\corrauth{guojh@ynao.ac.cn}
        }

   \institute{Yunnan Observatories, Chinese Academy of Sciences, Kunming 650216, China
   \and International Centre of Supernovae, Yunnan Key Laboratory of Supernova Research, Yunnan Observatories, Chinese Academy of Sciences, Kunming 650216, China
   \and University of Chinese Academy of Sciences, Beijing 100049, China
   \and School of Physics and Astronomy, Sun Yat-sen University, Zhuhai 519082, China.
   }

   \date{Received May 21, 2026}

 \abstract
   {Planet engulfment is a common phenomenon that leads to the accretion of planetary mass and the deposition of orbital angular momentum into the stellar envelope. As the final stage of star-planet interactions, this event is of great significance for understanding the post-engulfment evolution of stars, searching for observational evidence of such events, and investigating the complex interplay between stars and planets.}
   {This study focuses on the long-term variations in stellar metallicity and surface rotational velocity following planet engulfment, as well as the impact of rotationally induced mixing on the degree of stellar metal enrichment.}
   {We utilize the MESA code to explore how different stellar and planetary properties influence the metallicity and rotation of the host star after an engulfment event.}
   {Our models show that engulfment events during the later MS stage produce significantly higher enrichment than those at the ZAMS. While rotational mixing generally reduces the enrichment magnitude, metal-poor stars can maintain super-solar abundances for several Gyr, whereas high-metallicity stars are nearly insensitive to rotation. While planetary engulfment induces an instantaneous spin-up, this angular momentum injection is transient, and the star quickly re-converges to its nominal rotational evolution track. Therefore, the rotational acceleration during the MS cannot serve as a long-term criterion.}
   {The intensity and duration of stellar metal enrichment are heavily influenced by the thickness of the stellar convective envelope and their MS lifetime. While stars with longer MS lifetime tend to possess thicker convective envelope. Our research indicates that searching for evidence of planetary engulfment involves a critical trade-off between the long-lived but diluted signals in $0.7~M_{\odot}$ stars and the intense yet short-lived enrichment in $1.2~M_{\odot}$ stars.
   In conclusion, $1.0\,M_{\odot}$ stars which have relatively long MS lifetime and significant metal enrichment levels represent superior targets for observation.}

\maketitle
\nolinenumbers

\section{Introduction}\label{sec:Introduction}
Since 1995, over 6000 exoplanets have been discovered. Many of these have short orbital periods and are close proximity to their host stars \citep{Sanchis2014,2021ARA&A..59..291Z}. Such planets are highly susceptible to orbital decay due to tidal dissipation, potentially leading them to plunge into the convective envelopes of their host stars and alter the chemical abundances of the photosphere \citep{2024AJ....168..101D,2025A&A...693A..47S,2025AJ....170...93C}. The uncertain planetary occurrence rate and the transient nature of the engulfment events, hinders direct detection. Inference of these events therefore depends on identifying anomalous stellar properties.

\cite{1997ApJ...491L..51L} first modelled how planetary engulfment could enhance metallicity in F-type stars and examined its dependence on stellar mass. \cite{2009ApJ...704L..66M} compared the chemical composition of the Sun with 11 solar twins and found it to be deficient in refractory elements. Subsequent studies proposed various explanations for this anomaly \citep{2014A&A...564L..15A,2017AN....338..442A,2020MNRAS.493.5079B} . \cite{2015A&A...582L...6S} suggested that solar twins may exhibit overabundances of refractory elements due to the accretion of planetary material. \cite{2016ApJ...833L..24A,2020ApJ...897L..20A} linked planetary engulfment to \ce{^{7}Li} anomalies in red giants. \cite{2021AJ....162..273S} identified that the optimal conditions for \ce{^{7}Li} enrichment from planetary engulfment occur in $1.4$--$1.6 M_{\odot}$ stars at the main-sequence turnoff or early subgiant phase. \cite{2023MNRAS.518.5465B} carried out calculations for planetary engulfment by stars of different metallicities and found that the resulting chemical signature is more pronounced in stars with low Z. \cite{2023Natur.617...55D} reported that the ZTF SLRN-2020 event, characterized by an exceptionally low optical luminosity of $\sim 10^{35}\text{ erg s}^{-1}$ and long-lived infrared excess, marks the first direct observation of the engulfment of a low-mass giant planet ($\lesssim 10\,M_{\text{Jup}}$) by a Sun-like star. \cite{2026ApJ..1004..193K} confirms that the subgiant TOI-5882 exhibits significant lithium enrichment, suggesting that this anomaly likely results from the engulfment of a super-Earth or Neptune-mass planet. \cite{2026ApJ..1003...67L} simulated the process of $0.5\text{--}1.4\,M_{\odot}$ main-sequence stars engulfing terrestrial rocky planets and revealed that $1.0\text{--}1.4\,M_{\odot}$ stars are the most promising targets for detecting metal pollution (particularly Al, Ca, V, and Li), owing to their thin outer convective zones.

Open clusters and binary systems serve as excellent laboratories for detecting signatures of planetary engulfment. They formed simultaneously from the same molecular cloud and share identical compositions\citep{2017A&A...604L...4S,2018ApJ...854..138O}. Identifying evidence of planet engulfment in binary systems would significantly advance our understanding of planetary formation and evolution. \cite{2021NatAs...5.1163S} conducted a statistical study of 107 binary systems composed of Sun-like stars with similar effective temperatures and surface gravities. They found that 20–35\% of these stars exhibit anomalous lithium and iron abundances, suggesting planetary engulfment as the cause. \cite{Liu2024} report high-precision chemical abundances for a homogeneous sample of 91 co-natal stellar pairs and identified at least seven new cases of planetary engulfment.

When a star engulfs a planet, the planet transfers angular momentum and material to the star, affecting the star’s rotational velocity and chemical composition. Previous studies on planet engulfment have largely neglected the coupling between stellar rotation and rotational mixing. Some works have focused solely on rotational effects without accounting for changes in chemical abundances (e.g., \cite{2009ApJ...700..832C,2017MNRAS.465..149J,2016A&A...593A.128P,2020ApJ...889...45S,2020A&A...643A..34O, 2023RAA....23i5014G, 2024MNRAS.533.2199G}), while others have examined chemical enrichment in detail but ignored the role of rotation (e.g., \cite{2021NatAs...5.1163S, 2021AJ....162..273S, 2023A&A...670A.155C, 2023MNRAS.518.5465B, 2025A&A...693A..47S}). Therefore, it is essential to consistently couple chemical evolution with stellar rotation and how mixing processes influence post-accretion stellar structure and chemical compositions. We use \textbf{MESA} (Modules for Experiments in Stellar Astrophysics, \cite{2011ApJS..192....3P,2013ApJS..208....4P,2015ApJS..220...15P,2018ApJS..234...34P,2019ApJS..243...10P}) to explore the effects of varying stellar and planetary parameters.

This paper is structured as follows: Section \ref{sec:Stellar Model} is dedicated to our stellar evolution model using \textbf{MESA}, Section \ref{sec:Results} to the effects of accretion events on the star, Section \ref{sec:discussion} to a discussion of the results, Section \ref{sec:Conclusion} to a summary, and prospects for future work.

\section{Stellar Model}\label{sec:Stellar Model}
     \subsection{Stellar and accretion model}\label{subsec:stellar accretion model}
     We employ the open-source 1D stellar evolution code \textbf{MESA}  to compute the evolution of stars after accreting refractory material. Our stellar models are divided into non-rotating and rotating stars. In non-rotating stars, we include mixing mechanisms such as convection, overshoot, thermohaline mixing, and gravitational settling. In rotating stars, rotational mixing is enabled, with specific parameters following \cite{2000ApJ...544.1016H}, and we adopt the magnetic braking model from \cite{2015ApJ...799L..23M}, which will be detailed in the next subsection.

        \begin{figure}[!ht]
                      \centering
                      \includegraphics[width=0.5\textwidth]{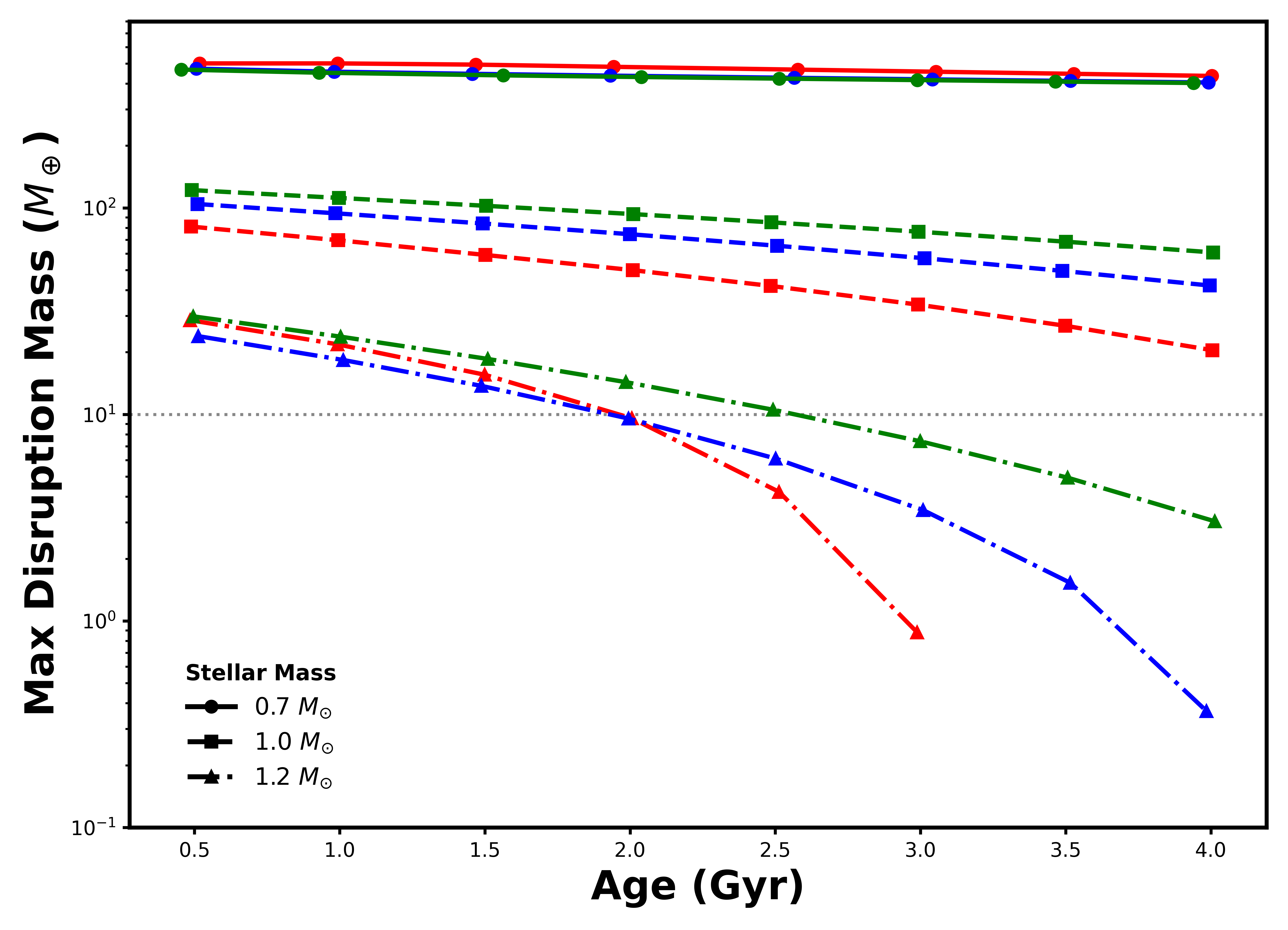}
                      \caption{The maximum masses of rocky planets that can undergo tidal disruption outside a host star are shown for different models across various evolutionary ages based on the Roche limit, utilizing the planet mass-radius relation from \cite{2007ApJ...659.1661F}. Different symbols denote models with varying stellar masses, while different colors represent different internal metallicities (red, blue, and green correspond to $\text Z_{int}$ = 0.005, 0.012, and 0.022, respectively).}
                      \label{fig:disruted_mass}
        \end{figure}

   According to Equation (\ref{eq:Teff_planet}), for a rocky planet with extremely close-in orbit, surface temperature can reach 3300~K (with semi-major axis equal to $1.44 R_*$ of a solar-like star, according to Equation (\ref{eq:roche_limit}) and \cite{2013ApJ...773L..15R}), exceeding the melting point of iron :

    \begin{equation}\label{eq:Teff_planet}
        T_{\text{eq}} = T_* \cdot \sqrt{\frac{R_*}{2a}} \cdot (1 - A_B)^{1/4}
    \end{equation}

    \begin{equation}\label{eq:roche_limit}
        d_{\rm roche} = 2.44 \cdot R_* \cdot\left( \frac{\rho_*}{\rho_{\rm p}} \right)^{1/3}
    \end{equation}

   where $T_*$ is the surface temperature of the star, $A_B$ is the Bond albedo of the planet (0.1 for rocky planets, \cite{2019Natur.573...87K}), $R_*$ is the stellar radius, $\rho_*$ and $\rho_p$ is the mean density of the star and planet, $a$ is the semi-major axis of the planet.

    The extreme high temperature renders the planet molten and departing from rigid-body characteristics \citep{2011Icar..213....1L,2016ApJ...828...80K}.
  Furthermore, \cite{2020ApJ...894....8P} shows that for an $8\,M_\oplus$ planet, the pressure at the core-mantle boundary reaches $10^{11}\text{ Pa}$, which is far greater than the strength of peridotite (tens of MPa, \cite{1999Tectp.303..175H}) and iron ($\sim 7\text{ GPa}$, \cite{CLATTERBUCK20032271}). Therefore, the effect of material rigidity for ultra-short-period rocky planets is only a minor correction, which does not invalidate the fluid hydrostatic equilibrium assumption. Their findings indicate that the material strength of rocky planets in close-in orbits is negligible, thus justifying the approximation of such planets as fluids. Under strong tidal forces, the planet undergoes significant prolate ("rugby-ball-shaped") tidal deformation, which amplifies its tidal response. Therefore, using the Roche limit (Equation (2))of the fluid to describe such planets is well justified \citep{2013ApJ...773L..15R,2017MNRAS.465..149J,2018ApJ...864..169J,2024AJ....168..101D}.

    Figure \ref{fig:disruted_mass} illustrates the maximum masses of rocky planets that can undergo tidal disruption outside host stars for different models at various evolutionary ages according to Equation (\ref{eq:roche_limit}). It can be seen that, except for the late evolutionary stages of the 1.2 $M_{\odot}$ model, all other models are capable of disrupting rocky planets with masses of 10 $M_{\oplus}$. Consequently, we assume that the planet undergoes tidal disruption before reaching the stellar surface. Given that MESA models mass gain exclusively via accretion \citep{2011ApJS..192....3P}, and the considering the huge disparity between the timescales of planetary disruption (Equation (\ref{eq:disruption time})) and stellar main-sequence evolution (Equation (\ref{eq:evolve time})) :

    \begin{equation}\label{eq:disruption time}
            \tau_{\text{disrp}} \sim \frac{1}{\sqrt{G\rho_p}}
    \end{equation}

    \begin{equation}\label{eq:evolve time}
            \tau_{\text{nuc}} \sim 10^{10} \times \left( \frac{M_*}{M_\odot} \right)^{-2.5} \text{years}
    \end{equation}

     we consider a simplified scenario of instantaneous accretion following \cite{2022MNRAS.516.3354S,2023MNRAS.518.5465B,2025A&A...693A..47S,2026arXiv260531060M}. To ensure numerical stability within MESA, we adopt a constant accretion rate of $10^{-9}\,M_{\odot}\,\text{yr}^{-1}$ .

   In rotating star models, angular momentum transfer from the accreted material is included. The orbital angular momentum of the accreted matter is deposited into the outermost layer of the star and subsequently transported inward through the stellar interior, as described in Section~\ref{subsection:Mixing processes}. To investigate the upper limit of stellar spin-up caused by planetary engulfment, We set the accreted angular momentum equal to the orbital angular momentum of the material, using \texttt{accreted\_material\_j} in MESA, where the specific angular momentum is given by
    \begin{equation}
        L = \sqrt{ G M_{*} a(1-e^2)}
    \end{equation}
    where $M_*$ stands for the mass of the star and $e$ is the eccentricity of the planetary orbit, respectively \citep{2026enap....1..538M}. And in this work, $e = 0$,  $a = 1.44 \cdot R_*$ ($1.44\cdot R_*$ is mainly disscussed in the result, the angular momentum at different locations is presented in Section \ref{subsec:locations}). Regarding the composition of the accreted material, we adopt the bulk Earth composition from the work of \cite{2003TrGeo...2..547M}.  For simplicity, no planetary energy transfer to the star is considered during the accretion process.

     \subsection{Magnetic braking}\label{subsec:Rotation and Magnetic Braking}
    In our stellar rotation model, we employ the magnetic braking formulation from \cite{2015ApJ...799L..23M}. During the pre-main sequence (PMS) phase, the star is forced to maintain at a constant angular velocity to simulate the effect of disk locking. The duration of the disk locking is set to be 3 Myr \citep{2021ApJ...912...65G}. The specifics of the magnetic braking are as follows:
    \begin{equation}
        \frac{dL_{\text{wind}}}{dt} =
        \begin{cases}
            -T_0 \left( \dfrac{\tau_{cz}}{\tau_{cz\odot}} \right)^p \left( \dfrac{\Omega_\ast}{\Omega_\odot} \right)^{p+1} & \text{(unsaturated)}, \\[3ex]
            -T_0 \chi^p \left( \dfrac{\Omega_\ast}{\Omega_\odot} \right) & \text{(saturated)},
        \end{cases}
        \label{eq:wind_loss}
    \end{equation}
    \noindent where the torque scaling coefficient $T_0$ and other parameters are defined as:
    \begin{align}
        T_0 &= K \left( \frac{R_\ast}{R_\odot} \right)^{3.1} \left( \frac{M_\ast}{M_\odot} \right)^{0.5} \gamma^{2m}, \label{eq:T0} \\[1.5ex]
        \gamma &= \sqrt{1 + (u/0.072)^2}, \label{eq:gamma} \\[1.5ex] 
        u &= \frac{\Omega_\ast}{\Omega_{\text{crit}}}, \quad \Omega_{\text{crit}} = \sqrt{\frac{GM_\ast}{R_\ast^3}}. \label{eq:u}
    \end{align}
     where \( L_{\text{wind}} \) is the angular momentum lost by the magnetic stellar wind, \(\Omega_{*}\), and \( \Omega_{\text{crit}} \) are the angular rotation rate and critical angular rotation rate, respectively. The constants \( K, m, p \), and \( \chi \) are free parameters. The saturated and unsaturated states  represent two different states of stellar magnetic activity, both of which have a strong correlation with the Rossby number \( R_0 = \frac{2\pi}{\Omega_{*}\tau_{cz}} \). The convective turnover timescale \( \tau_{cz} \) can be expressed as \cite{2021ApJ...912...65G}:
     \begin{equation}
         \tau_{cz}(r) = \alpha_{\text{MLT}}H_p(r)/v_c(r),
     \end{equation}
     where \( H_p(r) \) is the scale height, \( v_c(r) \) is the convective velocity at radius \( r \), and \( \alpha_{\text{MLT}} \) is the convective mixing length. \( \tau_{cz} \) is the turnover timescale of the convective zone, which is defined in the location where \( r = r_{\text{BCZ}} + 0.5H_p(r) \), and \( r_{\text{BCZ}} \) is the radius of the bottom of the outer convection zone. The value of \( \alpha_{\text{MLT}} \) is 1.82.

     \( R_0 \leq R_{\text{osat}} \) is the saturation region, and vice versa. The parameter \( \chi = R_0/R_{\text{osat}} \) relies on the critical value \( R_{\text{osat}} = 0.14 \) (Wright et al. 2018).
     The solar \( R_0 \) is around 2 (\cite{2016MNRAS.462.4442S}). Thus,we take \( \chi = 14.28 \). The constant \( m \) is set to 0.22. In present work, the values of \( K \) and \( p \) are set to \( 1.4 \times 10^{30} \) and 2.6, respectively.

     \subsection{Mixing processes}\label{subsection:Mixing processes}
        We have enabled rotational mixing and angular momentum transport in the rotating star model. In MESA\citep{2011ApJS..192....3P,2013ApJS..208....4P}, the chemical composition evolution is described by the diffusion equation:
        \begin{equation}
            \frac{\partial X_i}{\partial t} = \frac{\partial}{\partial m} \left[ (4\pi r^2 \rho)^2 D_{\text{mix}} \frac{\partial X_i}{\partial m} \right] + \left( \frac{\partial X_i}{\partial t} \right)_{\text{nuc}}
        \end{equation}
        where $D_{\text{mix}}$ represents the total mixing diffusion coefficient, which includes convective mixing, overshooting, thermohaline mixing($D_{\text{th}}$), and rotational mixing ($D_{\text{rot}}$):
        \begin{equation}
            D_{\text{mix}} = D_{\text{non-rot}} + D_{\text{rot}}
        \end{equation}
        Here, $D_{\text{non-rot}}$ is the sum of the diffusion coefficients for all mixing processes except rotational mixing.

        In stellar interiors, this instability is typically triggered by nuclear reactions (e.g., $^3\text{He}(^3\text{He},2p)^4\text{He}$, which reduces the local mean molecular weight) or by accretion of metal-rich material onto the stellar surface. The instability is quantified by the density ratio $R_0$ :
                \begin{equation}
                    R_0 = \frac{\nabla - \nabla_{\mathrm{ad}}}{(\phi/\delta)\nabla_\mu}
                \end{equation}
        where
        $\nabla \equiv \mathrm{d} \ln T / \mathrm{d} \ln P$ is the temperature gradient,
        $\nabla_{\mathrm{ad}}$ is the adiabatic temperature gradient,
        $\nabla_\mu \equiv \mathrm{d} \ln \mu / \mathrm{d} \ln P$ is the mean molecular weight gradient (positive when $\mu$ increases with pressure),
        $\phi \equiv (\partial \ln \rho / \partial \ln \mu)_{P,T}$ and $\delta \equiv -(\partial \ln \rho / \partial \ln T)_{P,\mu}$ are equation-of-state coefficients.

        The system is unstable to thermohaline mixing when $1 < R_0 < 1/\tau$. For $R_0 < 1$, it is unstable to Ledoux convection, while for $R_0 > 1/\tau$, it is stable, where $\tau$ is the Lewis number.

        The diffusion coefficient for thermohaline mixing is :
        \begin{equation}
            D_{\mathrm{th}} = C_t \frac{16\sigma T^3}{3\kappa\rho^2 C_P} \frac{\nabla_\mu}{\nabla_{\mathrm{rad}} - \nabla_{\mathrm{ad}}}
        \end{equation}
        where $C_t$ is a dimensionless efficiency parameter and $\sigma$ is the Stefan-Boltzmann constant.

        The total diffusion coefficient for rotational chemical mixing, $D_{\mathrm{rot}}$, is computed as the sum of contributions from multiple rotationally induced instabilities, scaled by an efficiency factor $f_c$:

        \begin{equation}
            D_{\mathrm{rot}} = f_c \sum_k D_k
        \end{equation}

        where the summation index $k$ runs over the following instabilities: Eddington–Sweet circulation (ES), dynamical and secular shear instabilities (DSI, SSI), Goldreich–Schubert–Fricke instability (GSF), and Solberg–Høiland instability (SH). Following \cite{2000ApJ...544.1016H}, we adopt $f_c = 0.033$  and set the chemical gradient sensitivity parameter $f_\mu = 0.05$ .

        Angular momentum transport is similarly treated using a diffusion approximation:
        \begin{equation}
        \begin{split}
        \left(\frac{\partial \omega}{\partial t}\right)_{m} = & \frac{1}{i}\left(\frac{\partial}{\partial m}\right)_{t} \left[ \left(4 \pi r^{2} \rho\right)^{2} i \nu \left(\frac{\partial \omega}{\partial m}\right)_{t} \right] \\
        & - \frac{2 \omega}{r} \left(\frac{\partial r}{\partial t}\right)_{m} \left(\frac{1}{2} \frac{d \ln i}{d \ln r}\right)
        \end{split}
        \end{equation}

        where $i$ is the specific moment of inertia of a shell at mass coordinate $m$, and $\nu$ is the turbulent viscosity determined as the sum of the diffusion coefficients for convection, double diffusion, overshooting and rotationally induced instabilities It is important to note that the diffusion coefficient $D$ for chemical mixing and the viscosity $\nu$ for angular momentum transport are not physically equivalent. Although turbulence transports both material and angular momentum, their efficiencies may differ significantly in the presence of strong mean molecular weight gradients ($\nabla_\mu$) or magnetic fields.

        \section{Results}\label{sec:Results}
        
        In this section, we compute an extensive grid of stellar parameters. The stellar mass range is from \(0.7\ M_{\odot}\) to \(1.2\,M_{\odot}\) and these stars evolve from PMS to the terminal-age main sequence (TAMS). We consider accretion events occurring at different evolutionary stages: zero-age main sequence (ZAMS), and mid-main sequence.
       \begin{figure*}[!ht]
            \centering
            \includegraphics[width=1.0\textwidth, trim={0 1.2cm 0 0}, clip]{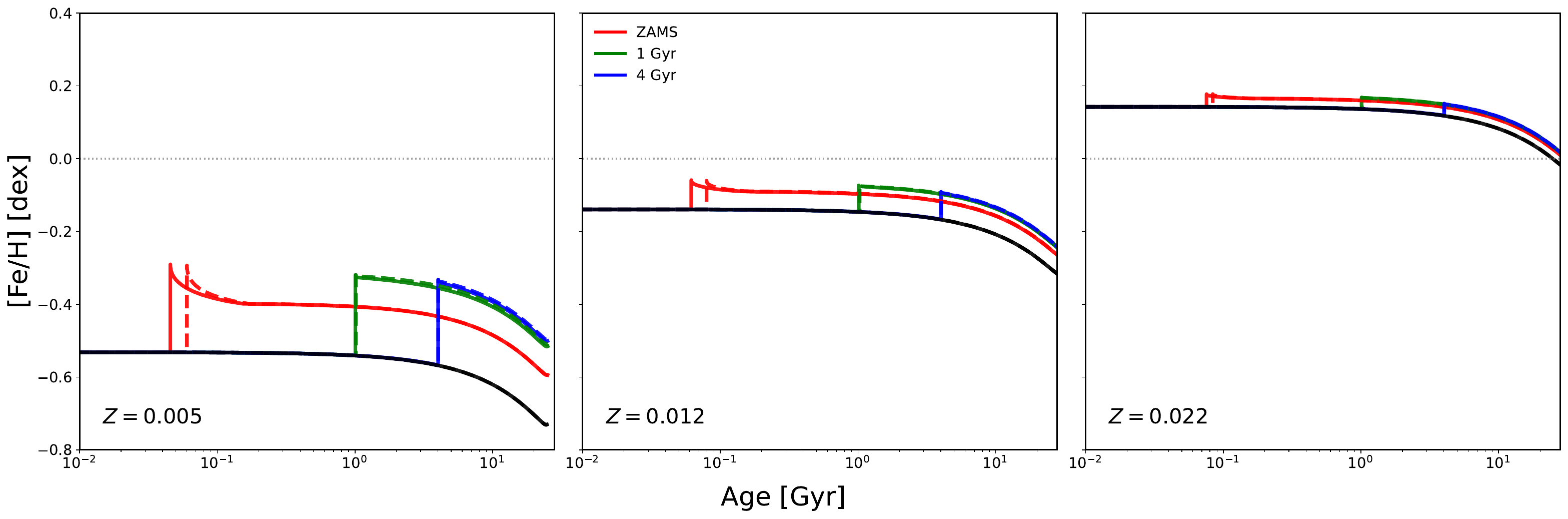}
            \includegraphics[width=1.0\textwidth, trim={0 1.2cm 0 0}, clip]{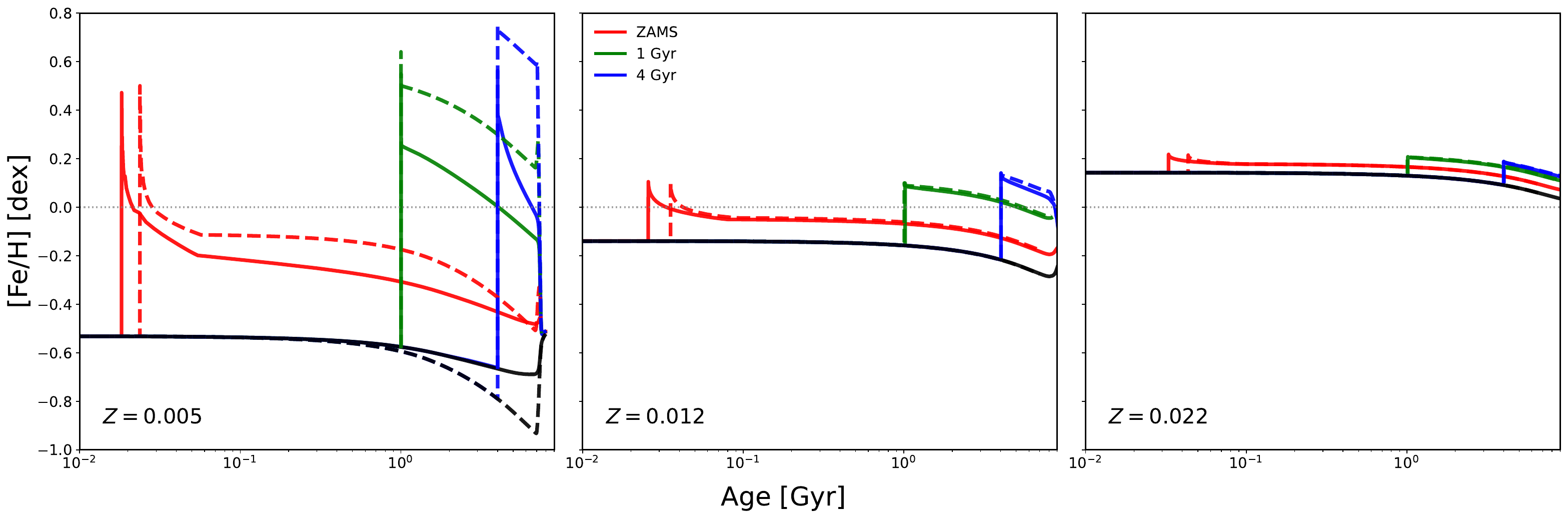}
            \includegraphics[width=1.0\textwidth]{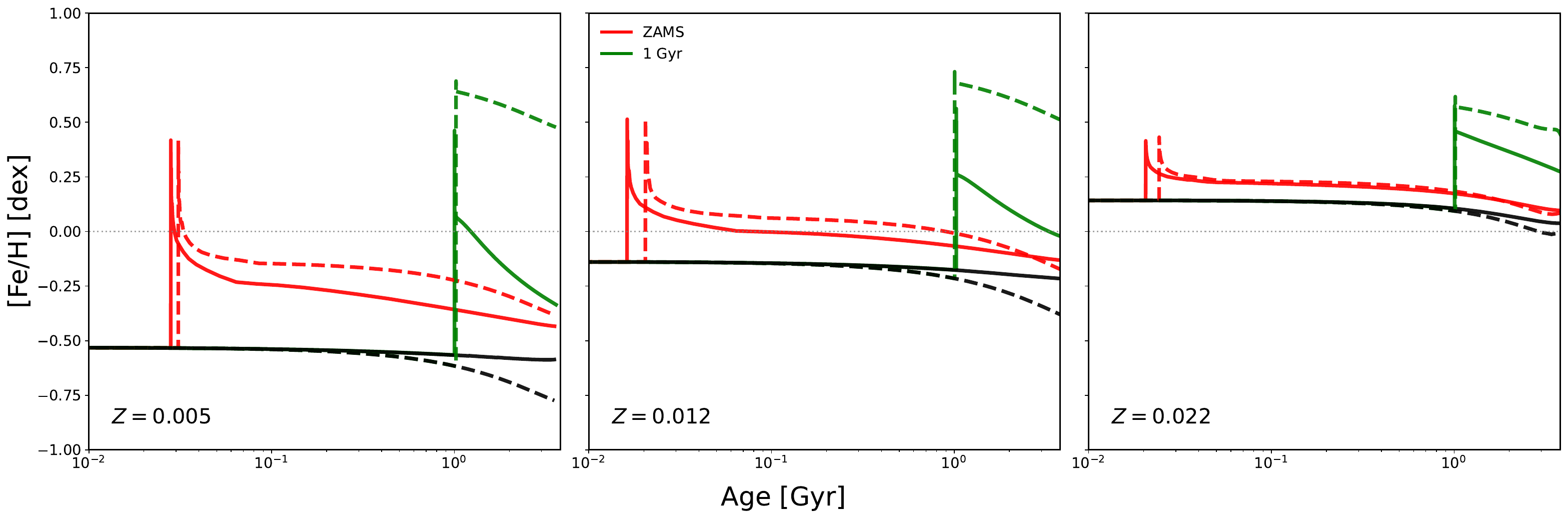}
        
            \caption{Evolution of [Fe/H] for $0.7\,M_{\odot}$ (top panel), $1.0\,M_{\odot}$ (middle panel), and $1.2\,M_{\odot}$ (bottom panel) stars with various initial metallicities following the engulfment of a $10\,M_{\oplus}$ planet at different epochs. The different colored lines denote the various times of the engulfment event. Solid lines represent models with rotation, while dashed lines represent non-rotating models. Black lines represent the non-accreting model. The gray dotted line represents the solar abundance. The initial period ($P_{\text{rot,init}}$) of the rotation model is 8 days.}
            \label{fig:time_combined}
        \end{figure*}
             \begin{figure*}[t]
                  \centering
                  \includegraphics[width=1.0\textwidth]{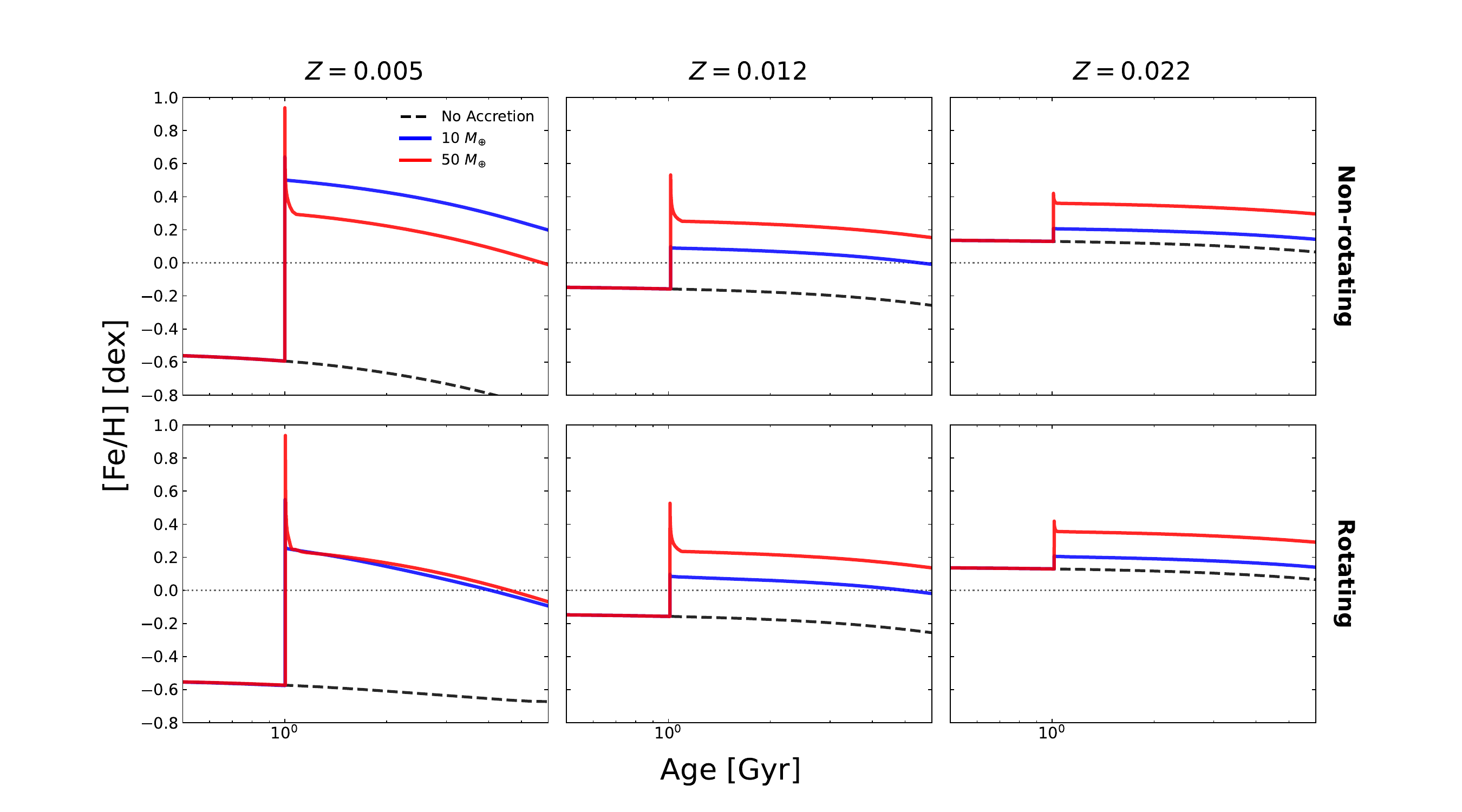}
                  \caption{Evolution of [Fe/H] following the engulfment of a $10\,M_{\oplus}$ (blue lines) and a $50\,M_{\oplus}$ (red lines) planet at 1 Gyr for 1 $M_\odot$ stars with various initial metallicities. Each columns corresponds to a different metallicity ($Z = 0.005, 0.012, \text{and } 0.022$), while the rows distinguish between the non-rotating case (upper panel) and the case accounting for rotation (lower panel, $P_{rot, init}$ = 8 days). The gray dotted line represents the solar abundance.}
                  \label{fig:mass_planet}
             \end{figure*}
        The hydrogen-rich gaseous envelope of a giant planet is susceptible to atmospheric escape driven by thermal energy deposited in the lower layers of the atmosphere of planet, intense XUV radiation and the tidal forces of the host star \citep{2024NatAs...8..920G}. Particularly as the planetary orbit continues to decay, the atmospheric escape rate driven by tidal forces and the planet's intrinsic thermal energy becomes so high \citep{2024NatAs...8..920G} that the envelope of a gas giant may be completely stripped away, leaving behind a ``remnant core" as exemplified by TOI-849b, CoRoT-7b and Kepler-10b \citep{2020Natur.583...39A, 2013MNRAS.433.3239K}. Therefore, the objective of this study is to investigate the evolutionary effects resulting from the engulfment of a $5 -50 M_{\oplus}$ planet composed of refractory heavy elements. \cite{2023MNRAS.518.5465B} calculates the chemical abundance variations in the stars with near-solar abundances ($Z = 0.012- 0.022$) following planetary engulfment. Planets also exist in the environments of certain metal-poor stars although the occurrence rate of gas giants around stars with low metallicity is relatively low \citep{2005ApJ...622.1102F}. Furthermore, the metal-poor stars possess thinner convective envelopes so they are expected to exhibit more pronounced surface enrichment, which provide superior candidates for identifying stars that have undergone engulfment events. Thus, our calculations cover the metallicity range of $Z = 0.005 - 0.022$.
        As mentioned above, rotation can introduce additional mixing effects. For the sake of comparison, we calculated both rotational and non-rotational scenarios for each example.

        In this section, we will separately discuss the impact of stellar mass (\ref{subsec:metallicities}), the timing of the engulfment event (\ref{subsection:time}), planetary mass (\ref{subsec:planet mass}), the initial stellar rotation period (\ref{subsec:initial period}) on the stellar metallicity enrichment following a planetary engulfment event, and how the duration of metal enrichment scales with all these parameters (\ref{subsec:duration}).

\subsection{Different stellar masses }\label{subsec:metallicities}
            Figure \ref{fig:time_combined} illustrates the influence of the time of engulfment on the $\rm [Fe/H]$ for stars with different masses and metallicities, comparing scenarios with and without rotation (10 $M_\oplus$ for the accretion model). The models simulate three distinct accretion epochs ranging from the ZAMS to an age of 4 Gyr (except the 1.2$M_\odot$ models because their MS lifetime cannot last for 4 Gyr). It is worth noting that rotating models reach the ZAMS earlier than their non-rotating counterparts due to the modification of the stellar structure by the rotation and the enforced disk-locking during the PMS.

            The red lines of Figure \ref{fig:time_combined} illustrate the evolution of [Fe/H] for stars of various masses and initial metallicities after accreting 10 $M_{\oplus}$ of material at the ZAMS. It is observed that stars with $1.0 \, M_{\odot}$ and $1.2 \, M_{\odot}$ exhibit higher levels of metal enrichment compared to the $0.7 \, M_{\odot}$ star. This is because the $0.7 \, M_{\odot}$ star possesses a deeper convective envelope than the stars with higher mass, leading to a stronger dilution of the accreted material within the convection zone. Consequently, $1.0 \, M_{\odot}$ and $1.2 \, M_{\odot}$ stars show more pronounced enrichment.

            The impact of rotation on surface abundances is evident for the stars with higher-mass and lower metallicity. In the non-accretion cases (black solid and dashed lines in Figure \ref{fig:time_combined}), for the $0.7\,M_{\odot}$ models and the high-metallicity $1.0\,M_{\odot}$ models, convection remains the dominant mixing mechanism (see last row of Figure \ref{fig:appendenx}). Consequently, the impact of rotation on the chemical evolution of these models is negligible (see first row, the middle and right pannels of second row in Figure \ref{fig:time_combined}). However, for stars with thinner envelope, in the absence of accretion (black lines in the left panel of the second row and all panels of the third row in Figure~\ref{fig:time_combined}), the models exhibit a gradual decline in surface iron abundance due to gravitational settling, and this depletion is more pronounced in non-rotating models (black dashed lines). In contrast, rotating models (black solid lines) maintain higher surface metallicities because rotational mixing partially counteracts atomic diffusion, effectively replenishing the photosphere with internal metals.

            The effect of rotation is more prominent when the accretion occurs for the cases with high mass and low metallicity (colored lines in the left panel of the second row and all panels of the third row in Figure \ref{fig:time_combined}). Generally, the surface $\rm [Fe/H]$ rises sharply when the star accretes metal-rich material. Subsequently, the accreted metals can be transported downwards and diluted into the interior, which proceeds more rapidly in the rotating model attributed to the rotation-induced mixing (colored solid lines). In non-rotating stars (colored dashed lines), the removal of accreted material only relies on the gravitational settling and thermohaline mixing. As a result, the enrichment of metallicity in non-rotating stars is more prominent. For instance, in the 1 Gyr accretion scenario, the metallicity enrichment ($\rm [Fe/H] > 0$) in the 1.0$\,M_{\odot}$ non-rotating model (green dashed lines) persists for nearly the entire remainder of the MS, whereas its rotating counterpart (green solid lines) decays to sub-solar values in 3 Gyr. Nevertheless, the stars with low metallicity that undergo late-stage engulfment events can maintain their metallicity above solar levels for several Gyr. As a result, they ``masquerade" as metal-rich stars, effectively masking their true initial metallicity. Comparing the cases with different metallicity, it is evident that the rotational dilution of accreted material is more pronounced in stars with lower metallicity ($Z = 0.005$). In fact, the stars with higher metallicity possess deeper convective envelopes, which trigger efficient magnetic braking. This leads to a rapid spindown of the angular velocity (see Figure \ref{fig:initial_period1}), subsequently reducing the efficiency of rotational mixing \citep{2010A&A...519A.116E}.

    \subsection{Different time of engulfment}\label{subsection:time}
              We now turn to the effects of engulfment timing on the evolution of surface abundance.
              We find that the level of metallicity enrichment increases as the timing of the engulfment event delayed (Figure~\ref{fig:time_combined}). To elucidate the physical basis for the results, we examine the internal mixing processes. When a star accretes material, vigorous convection rapidly homogenizes those material within the convective envelope at the first time. This process establishes an inverse mean molecular weight gradient ($\nabla_\mu$) at the envelope boundary between the convective and radiative zones, which drives the thermohaline mixing and transports heavy elements into the stellar interior (see Figure \ref{fig:discussion2}).

             The efficiency of thermohaline mixing is further modulated by the star's evolutionary stage. As the star evolves, gravitational settling transports more surface heavy elements inward, increasing the interior mean molecular weight. Consequently, planet engulfment at the ZAMS triggers efficient thermohaline mixing at the base of the convective zone. In contrast, for engulfment occurring at the later stage of MS, the pre-existing interior $\mu$-gradient hinders the establishment of an inverse molecular weight gradient (see the first row of Figure\ref{fig:appendenx}). This suppresses thermohaline mixing and allows a larger fraction of the accreted material to be retained in the surface layers \citep{2012ApJ...744..123T}. Furthermore, in the early stages near the ZAMS, the convective envelope may exhibit a greater depth compared to that of the subsequent MS phase, due to the onset of stable core hydrogen burning. Therefore, any accretion occurring near the ZAMS results in more vigorous convective mixing of the accreted material.

    \subsection{Different planetary masses}\label{subsec:planet mass}
           To investigate the influence of accreted mass on changes in stellar metallicity, we show two cases with accreted masses of 10 and 50 $M_\odot$(Figure \ref{fig:mass_planet}) at 1 Gyr. It is worth noting that, according to Figure \ref{fig:disruted_mass}, not all stars are capable of disrupting a $50~M_{\oplus}$ planet at their exterior. Nevertheless, stars of 0.7 and 1.0 $M_{\odot}$ are still capable of engulfing these $50~M_{\oplus}$ planets. We therefore include this case in our calculations. To isolate and examine the influence of other physical parameters on metallicity enrichment, we focus our subsequent discussion on the representative $1.0\ M_{\odot}$ model. We find that a larger accreted mass generally leads to greater metallicity enrichment, with the exception of the non-rotating models at low metallicity (Z=0.005). For the remaining cases, neither the rotating nor the non-rotating models exhibit the ``reversal" phenomenon. However, for the Z=0.005 case, the metallicity increase with 50$M_\oplus$ is not significantly higher than with 10$M_\oplus$. The phenomenon will be discussed in Section \ref{subsec:reversal}.
           In the $Z = 0.012$ and 0.022 models, the enrichment remains at super-solar levels for an extended period, independent of rotational effects. As discussed in Section \ref{subsection:time}, the mixing effects induced by rotation are not significant for stars with higher metallicities.

    \subsection{Different initial rotation periods}\label{subsec:initial period}
            \begin{figure*}[t]
               \centering
              \includegraphics[width=1.0\textwidth]{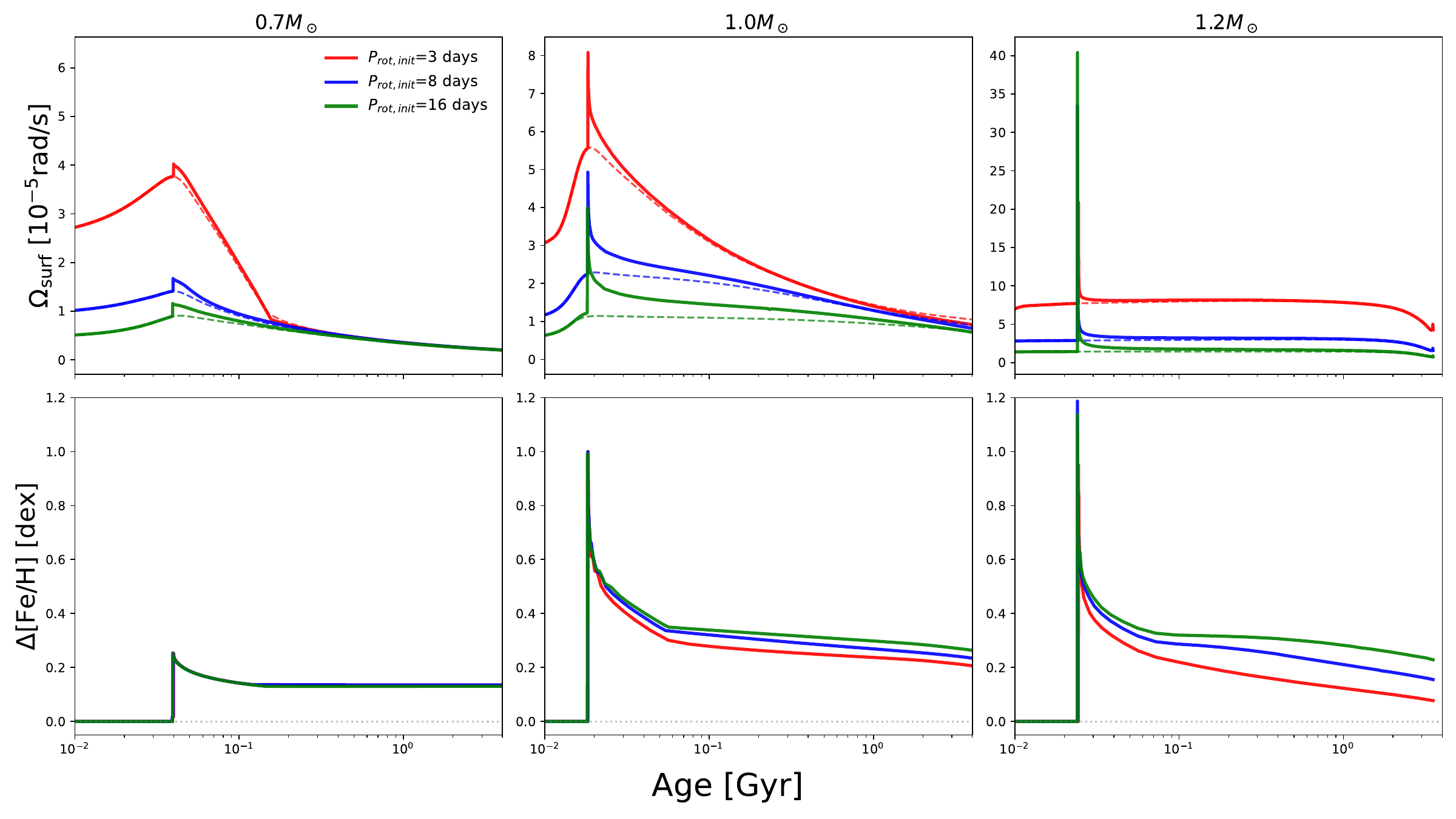}
              \caption{Upper panel shows the time evolution of surface angular velocity for stars with different masses(0.7 $M_\odot$,1.0 $M_\odot$,1.2 $M_\odot$) and initial rotation periods (3days, 8days, 16days). Solid lines represent models with accretion of $10\ M_{\oplus}$, while dashed lines represent non-accreting models. Lower panel shows the evolution of the metallicity difference between accreting and non-accreting stars. Their initial metallicity is $Z = 0.005$.}
              \label{fig:initial_period1}
            \end{figure*}

            \begin{figure}[t]
               \centering
              \includegraphics[width=0.5\textwidth]{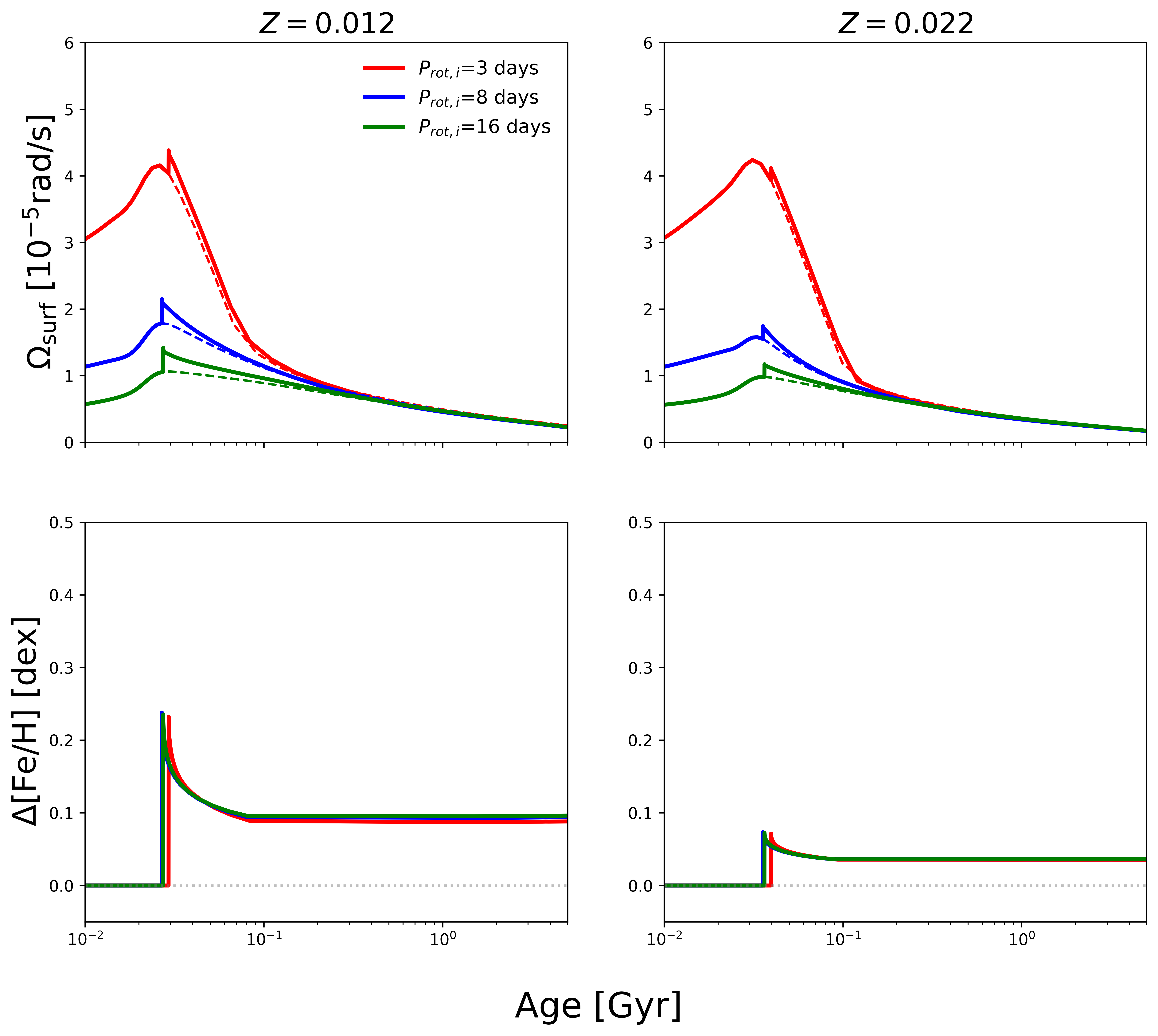}
              \caption{Similar to Figure \ref{fig:initial_period1}, but with columns corresponding to 1.0$M_\odot$ with different initial metallicities ($Z = 0.012, 0.022$).}
              \label{fig:initial_period2}
            \end{figure}
    \begin{figure*}[t]
            \sidecaption
            \includegraphics[width=12cm]{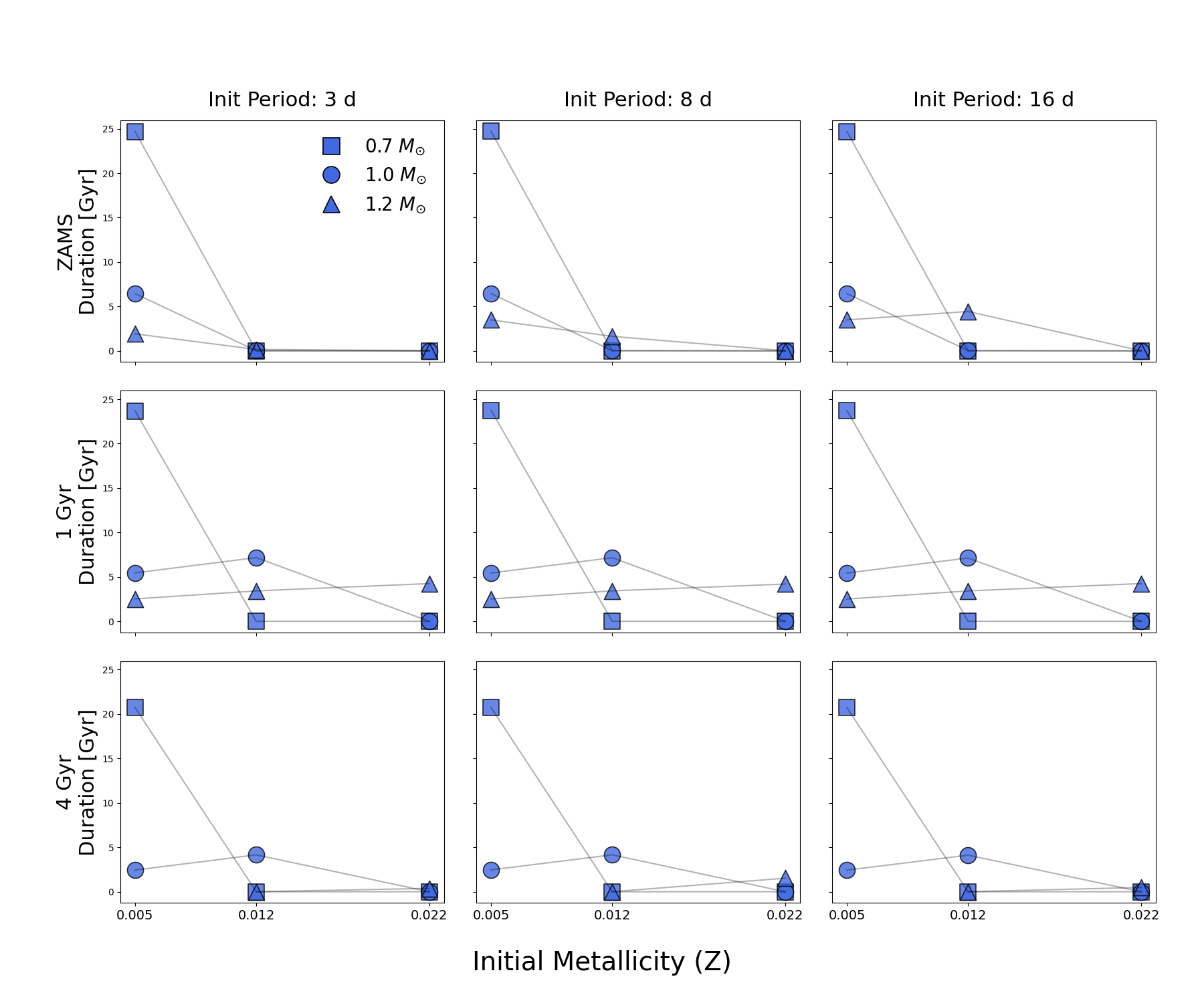}
            \caption{Duration of surface metallicity enrichment following the accretion of $10~M_{\oplus}$ of planetary material (except for the $1.2\,M_\odot$ models with metallicities of 0.012 and 0.022 undergoing engulfment at 4 Gyr, the accreted masses for these models are shown in Figure \ref{fig:disruted_mass}). The y-axis represents the duration over which the metal enrichment of the accreted model exceeds 0.1 dex compared to the non-accreted model. The x-axis represents the initial stellar metallicity ($Z_{\text{init}} = 0.005, 0.012, \text{and } 0.022$). Rows correspond to the timing of the accretion event: at the ZAMS, 1 Gyr, and 4 Gyr. Columns correspond to the initial stellar rotation periods: 3, 8, and 16 days. Different symbols indicate stellar masses: $0.7~M_{\odot}$ (squares), $1.0~M_{\odot}$ (circles), and $1.2~M_{\odot}$ (triangles). }
            \label{fig:duration1}
        \end{figure*}
        
\begin{figure*}[]
            \centering
            \includegraphics[width=0.85\textwidth]{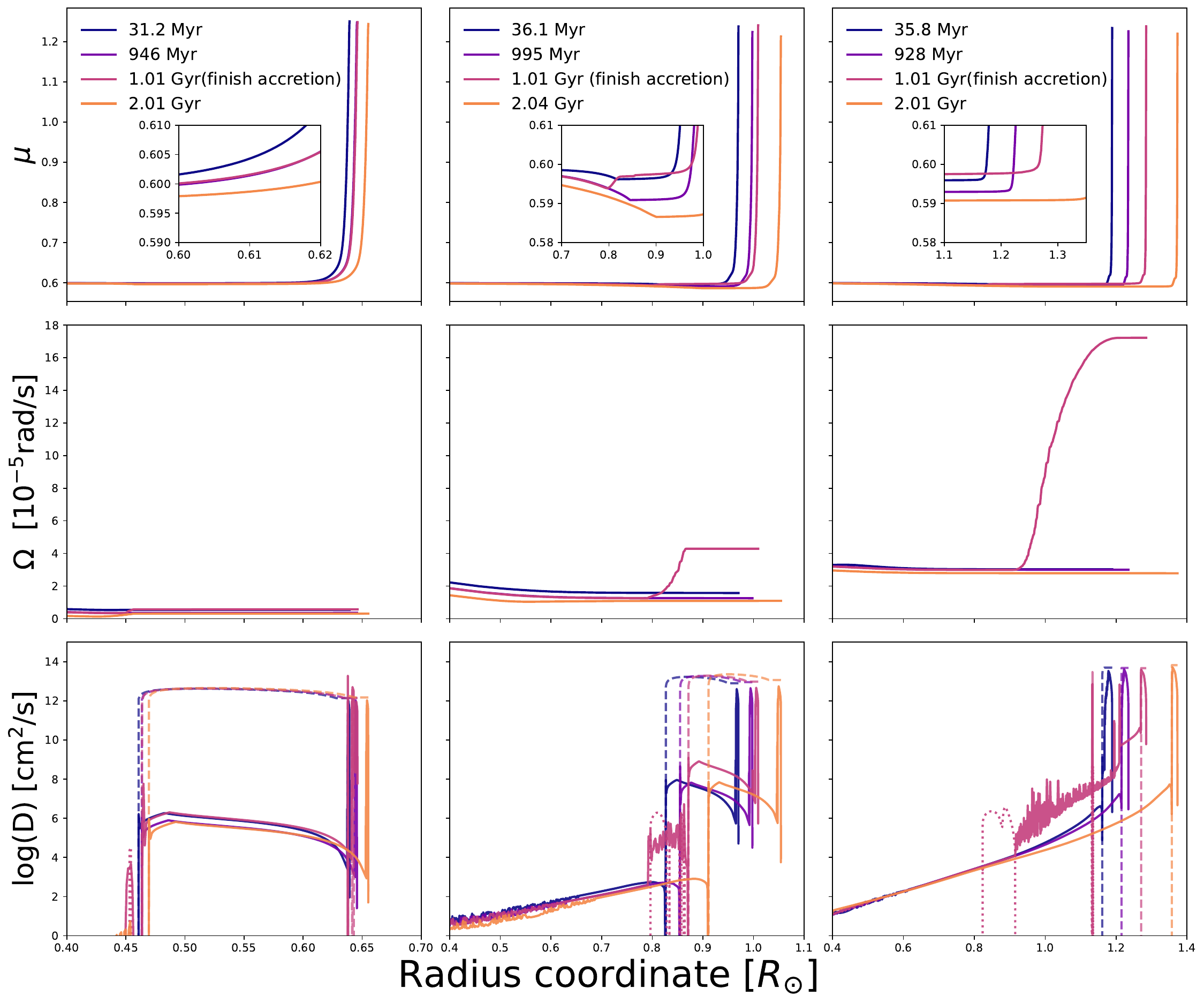}
            \caption{Spacial evolution of the internal structure and chemical abundance for $0.7 M_\odot$ (left column), $1.0 M_\odot$ (middle column) and $1.2 M_\odot$ (right column) stars with Z = 0.005 following a planetary engulfment event.
            Row 1: Profiles of the mean molecular weight ($\mu$). Time in the legend is star age. The insets provide a magnified view of the surface regions.
            Row 2: Angular velocity profiles ($\Omega$), illustrating the spin-up effect due to accretion.
            Row 3: Logarithm of the diffusion coefficients ($D$). Solid, dashed, and dotted lines correspond to mixing due to rotation, convection, and thermohaline instability, respectively. }
            \label{fig:appendenx}
          \end{figure*}

            Figure~\ref{fig:initial_period1} shows the evolution of angular velocity and $\Delta \mathrm{[Fe/H]}$ as a function of stellar age for stars which engulf a 10 $M_\oplus$ planet at ZAMS with varying initial rotation periods, where $\Delta \mathrm{[Fe/H]}$ represents the difference in metallicity between the accreted and non-accreted models. ($\Delta \mathrm{[Fe/H]}$ = $\mathrm{[Fe/H]_{accretion}}$ - $\mathrm{[Fe/H]_{non-accretion}}$). For low-mass stars ($0.7 M_{\odot}$), the evolutionary tracks of $\Delta \mathrm{[Fe/H]}$ exhibit a remarkable insensitivity to the initial rotation period. Rotational mixing is secondary to the dominant convective mixing in these stars with deep envelopes (see last row of Figure\ref{fig:appendenx}). As a result, accreted material is diluted homogeneously by convective mixing, leading to indistinguishable $\Delta \mathrm{[Fe/H]}$ decay profiles across different initial rotation periods.
            And as illustrated in the angular velocity evolution, although the models exhibit a dispersion in initial angular velocity of more than one orders of magnitude, ranging from $\sim 4.5 \times 10^{-6}$ rad/s to $2.4 \times 10^{-5}$ rad/s, corresponding to $P_{\mathrm{rot,init}}$ of 16 days and 3 days, respectively. Stellar wind efficiently extracts angular momentum via magnetic braking, causing these stars experience the rapid angular momentum loss. Consequently, by an age of $\sim 0.2$ Gyr, the angular velocity for all models converge significantly, uniformly decaying to the level of $10^{-6}$ rad/s.
            This phenomenon is also confirmed in metal-rich models of $1.0 M_{\odot}$. For stars with higher metallicities (Figure \ref{fig:initial_period2}), the thicker convective envelope leads to more efficient magnetic braking, resulting in negligible changes in angular velocity following planetary engulfment. Furthermore, as discussed in the preceding sections, the modulation of post-engulfment metal enrichment by rotational mixing is remarkably weak. Consequently, $\Delta \mathrm{[Fe/H]}$ does not exhibit significant dispersion relative to variations in the initial rotation periods.

            In contrast, as the stellar mass increases to $1.2 M_{\odot}$, the convective envelope thins significantly. In this regime, the efficiency of magnetic braking diminishes, which allows the stars to maintain long-term differences in angular velocity based on their initial conditions. Fast-rotating models ($P_{\mathrm{rot,init}}=3$ days) maintain higher angular velocities and rotation-induced mixing replace convection as the dominant role of material transport after accreting material (see last row in Figure \ref{fig:appendenx}). This accelerates the transportation of surface metal-rich material into the deep interior, resulting in a faster decay of $\Delta \mathrm{[Fe/H]}$. In contrast, slow rotators ($P_{\mathrm{rot,init}}=16$ days) retain surface abundance anomalies for a higher level. Therefore, in the higher-mass regime, the initial rotation period acts as a critical parameter and determines the evolutionary path of surface chemical abundances.

     \subsection{Metallicity enrichment duration}\label{subsec:duration}
          In this section, we examine a comprehensive parameter grid to assess the detectability of planetary engulfment. Current observational capabilities allow for the detection of metallicity variations as small as 0.1 dex \citep{2019ARA&A..57..571J}. Figure \ref{fig:duration1} illustrates the duration during which the surface metallicity enhancement remains above $0.1~\text{dex}$ following the accretion of $10~M_{\oplus}$ of planetary material, as a function of various stellar parameters. For metal-rich stars ($Z = 0.022$) with masses of $0.7$ and $1.0~M_{\odot}$, the enrichment level fails to reach this detectable threshold regardless of the accretion time. In contrast, for $1.2~M_{\odot}$ stars of same metallicity, the convective envelope remains sufficiently shallow to allow surface enrichment to exceed $0.1~\text{dex}$, except for the cases of the accretion on the ZAMS. As metallicity decreases, the enrichment becomes increasingly pronounced. For the cases with a metallicity of 0.012, even low-mass stars exhibit enrichment. At the lowest metallicities ($Z = 0.005$), all models show metal enrichment, and the lower the stellar mass, the longer the enrichment lasts. In some cases, persisting until the end of the main sequence. Increasing the accreted mass to $50~M_{\oplus}$ results in a sustained $0.1~\text{dex}$ enrichment across nearly all models, with the exception of the $0.7~M_{\odot}$ metal-rich model following ZAMS engulfment (Figure \ref{fig:duration2}). While the $1.2~M_{\odot}$ models exhibit high enrichment, their total observable duration is constrained by their shorter MS lifetimes (approximately $4\text{--}5~\text{Gyr}$), and the $1.2~M_{\odot}$ ($Z = 0.005$) model is absent from the final grid row as it does not reach a MS age of $4~\text{Gyr}$. Next, we increase the threshold for metallicity enrichment to 0.3 dex (Figure \ref{fig:duration3}). Under this condition, we find that none of the $0.7\ M_{\odot}$ models reach this threshold, whereas the $1\ M_{\odot}$ star exhibits the most persistent enrichment duration in most cases. Moreover, in the 1 Gyr engulfment scenario with $Z = 0.012$ and $0.022$, the sufficiently thin convective envelope of the $1.2\ M_{\odot}$ star allows it to achieve 0.3 dex, with its slowly rotating models sustaining this enrichment for 1–3 Gyr. Conversely, the $1.0\ M_{\odot}$ models fail to exceed this threshold. For the 4 Gyr engulfment scenario, 1.0$\ M_{\odot}$ and $1.2\ M_{\odot}$ stars with solar-like metallicity exhibit significant metal enrichment. In contrast, the enrichment in metal-rich $1.2\ M_{\odot}$ stars remains below 0.3 dex.

          These findings underscore that the long-term visibility of accretion signatures is a delicate balance between signal intensity and stellar longevity. Ultimately, the intensity of metal enrichment is dictated by the thickness of the stellar convective envelope, whereas its duration is governed by the star’s MS lifetime. Because stellar evolution remains stable during the MS and intense mixing occurs only briefly after accretion, a signal that survives the initial mixing phase is likely to remain detectable until the end of the MS. However, a critical trade-off exists where stars with longer MS lifetimes, such as low-mass or metal-rich stars, typically possess deeper convective envelopes that dilute the accreted material. For instance, while a metal-poor $0.7~M_{\odot}$ star can sustain a signal for tens of Gyr, its enrichment only marginally exceeds observational precision. Conversely, a $1.2~M_{\odot}$ star can surpass solar abundance levels but only for approximately $3~\text{Gyr}$. This interplay defines the effective observation window for engulfment events. In this context, a $1.0~M_{\odot}$ metal-poor star offers the optimal compromise, providing high enrichment intensity alongside a substantial MS lifetime ($\sim7~\text{Gyr}$), making it the premier target for investigating the engulfment history and dynamical evolution of planetary systems.

\section{Discussion}\label{sec:discussion}
    \subsection{Structure evolution}\label{subsec:Structure Evolution}
\begin{figure}
            \centering
            \includegraphics[width=0.5\textwidth]{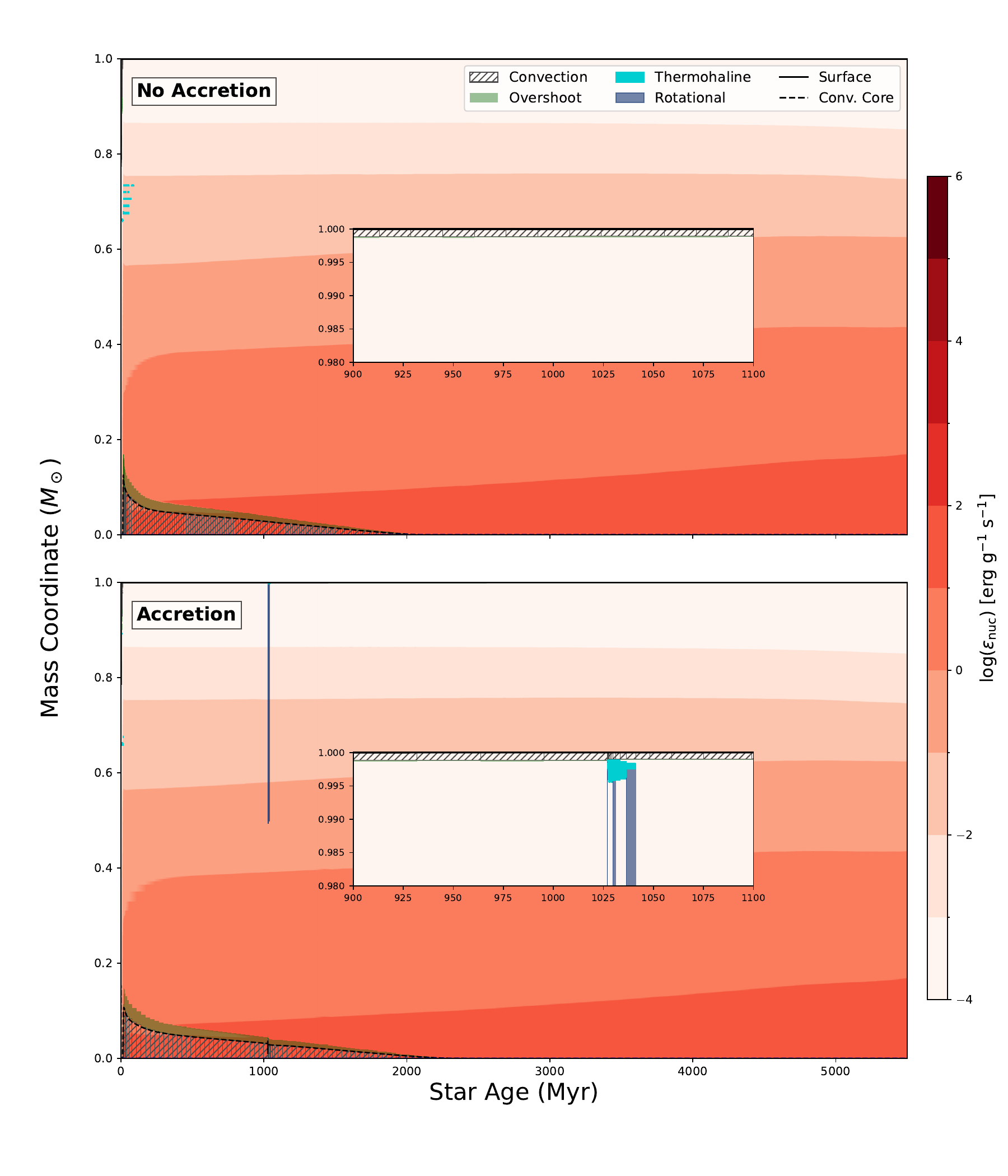}
            \caption{Kippenhahn diagrams illustrating the internal structural evolution of a $1.0~M_{\odot}$ star with an initial metallicity of $Z = 0.005$ and an initial rotation period of 8 days. The top panel shows the case without accretion, while the bottom panel depicts the results following the accretion of $10~M_{\oplus}$ material at an age of 1 Gyr. The background color contours represent the nuclear energy generation rate, $\log(\epsilon_{\rm nuc})$ [erg g$^{-1}$ s$^{-1}$], as indicated by the color bar on the right. Mixing regions are identified by different hatching and shading patterns: grey hatched areas (////) denote convective zones; green shaded regions indicate overshooting; cyan areas represent thermohaline mixing; and dark blue shaded regions indicate rotational mixing. Inset: Magnified view of the stellar surface layers ($0.98 \le M_r/M{_*} \le 1.0$)}
            \label{fig:discussion2}
        \end{figure}

        Next, we will discuss the evolution of the internal stellar structure. As illustrated in Figure \ref{fig:discussion2}, the stellar convective envelope undergoes a slight contraction following the cessation of accretion, a result that is consistent with the findings mentioned in the Section 3.1 of \cite{2023MNRAS.518.5465B}. Thermohaline and rotational mixing appear only during the accretion process and for a very brief period immediately afterward. Thermohaline mixing is confined to the vicinity of the convective boundary, whereas the influence of rotational mixing can extend down to a mass coordinate of approximately $0.5~M_\odot$. Because the accreted material consists entirely of metals, its incorporation into the stellar interior increases the opacity, leading to a transient adjustment of the convective core. However, the stellar structure promptly adjusts to this change, subsequently returning to a state nearly identical to its standard evolutionary track.

        \begin{figure}
              \centering
              \includegraphics[width=0.5\textwidth]{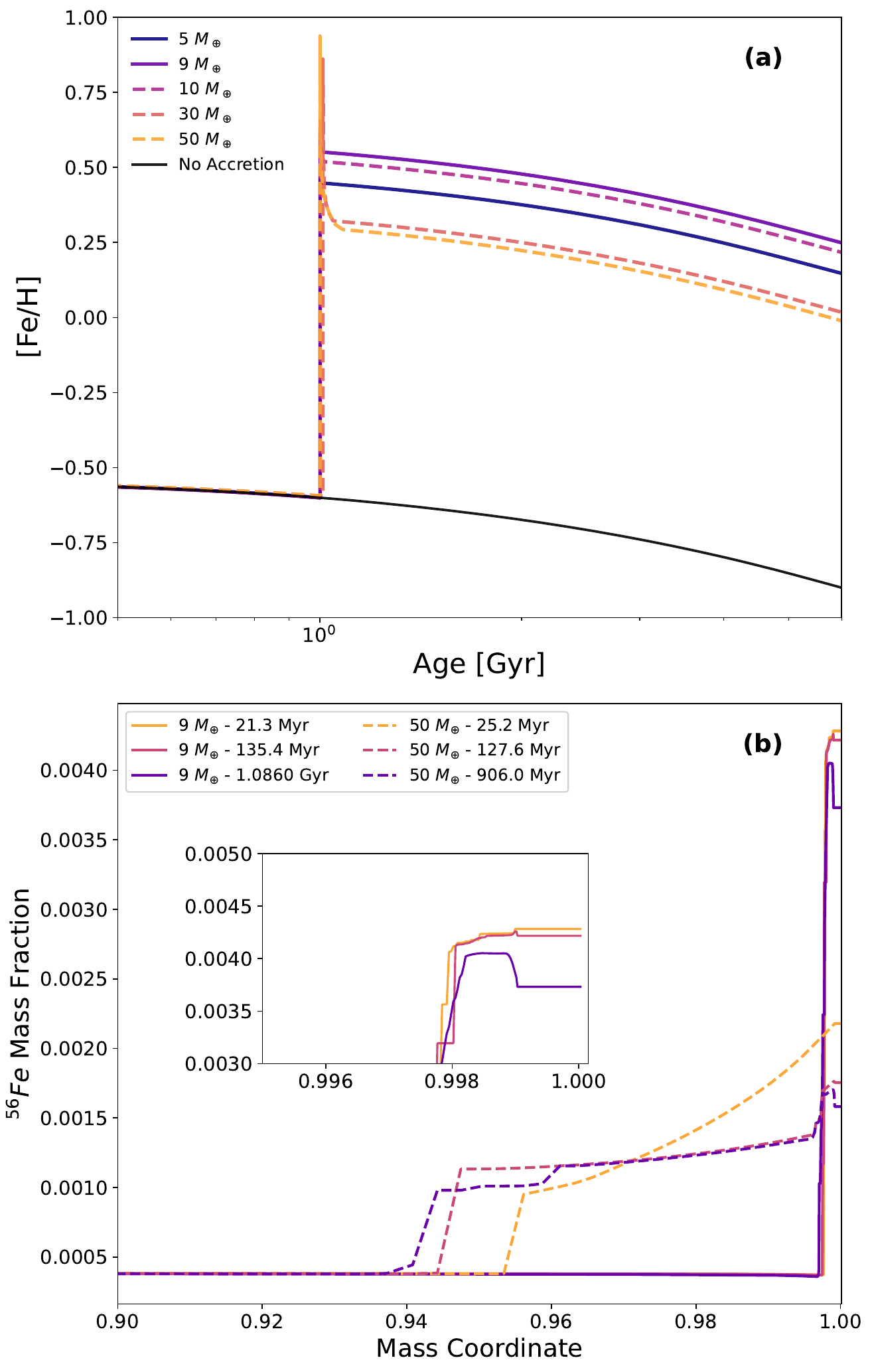}
              \caption{(a): The metallicity evolution of a 1 $M_\odot$ non-rotating star with initial metallicity $Z = 0.005$. The star accretes planetary material with masses ranging from 5 to 50 $M_\oplus$ at 1 Gyr. The ``reversal" phenomenon occurs when the star accretes planetary material of 9 $M_\oplus$. Dashed lines represent models with an accreted mass greater than 9 $M_\oplus$, yet exhibiting a lower level of metal enrichment compared to models that have accreted 9 $M_\oplus$. (b): The variation of the \ce{^{56}Fe} mass fraction in a star with stellar mass coordinates after accretion. The time in the legend represents the stellar age minus the time when accretion is completed (post-engulfment time). The dashed line corresponds to a star that has accreted 50 $M_\oplus$ material, while the solid line corresponds to a star that has accreted 9 $M_\oplus$ material. Inset: Magnified view of the stellar surface layers ($0.995 \le M_r/M{_*} \le 1.0$) of the 9 $M_\oplus$ accretion model.}
              \label{fig:1-50}
         \end{figure}

       The deposition of orbital angular momentum leads to a pronounced spin-up of the stellar envelope, creating a strong shear layer at the base of the convective zone. The effect is even more significant for $1.2 M_{\odot}$ stars (see the second row in Figure \ref{fig:appendenx}), because they possess smaller convective envelopes, the same input of angular momentum results in a more rapid acceleration of their angular velocity. However, while planetary engulfment induces an instantaneous spin-up, this angular momentum injection is transient, and the star quickly re-converges to its nominal rotational evolution track. Therefore, the rotational acceleration during the MS cannot serve as a long-term criterion.

    \subsection{Reversal phenomenon of metal enrichment in non-rotational cases}\label{subsec:reversal}
            To further investigate the ``reversal" phenomenon mentioned in Section \ref{subsection:time}, we modeled accreted masses ranging from 5 to 50 $M_\oplus$ (Figure~\ref{fig:1-50}(a)). We discovered that stellar metal enhancement reaches a maximum when the accreted mass is around 9 $M_\oplus$. When the accreted mass exceeds this value, the degree of metal enhancement decreases. For a star accreting 50 $M_\oplus$, although its surface \ce{^{56}Fe} abundance is higher immediately after accretion compared to a star accreting 9 $M_\oplus$, the \ce{^{56}Fe} is rapidly transported into the stellar interior (Figure~\ref{fig:1-50}(b)). In contrast, the \ce{^{56}Fe} in the star accreting 9 $M_\oplus$ remains near the surface.

            The peculiar ``reversal" at 1 Gyr suggests a competition between the accretion-induced inverse $s\mu$-gradient and the positive $\mu$-gradient built by gravitational settling (as discussed in Section \ref{subsection:time}). At this intermediate age, the internal positive $\mu$-gradient is established but remains fragile. While 9 $M_{\oplus}$ of accreted material is insufficient to overcome this barrier, a star with a high mass triggers a critical instability, leading to runaway thermohaline mixing that depletes surface metals more efficiently than in lower-mass accretion cases.

            Furthermore, rotation introduces a complex interplay with these mixing processes. As shown in the first column of Figure~\ref{fig:mass_planet}, for $10\,M_\oplus$ accretion, the rotating model shows lower enrichment ($0.2$ dex at 2 Gyr) than the non-rotating model ($0.4$ dex).
            \begin{figure*}
              \sidecaption
              \includegraphics[width=12cm]{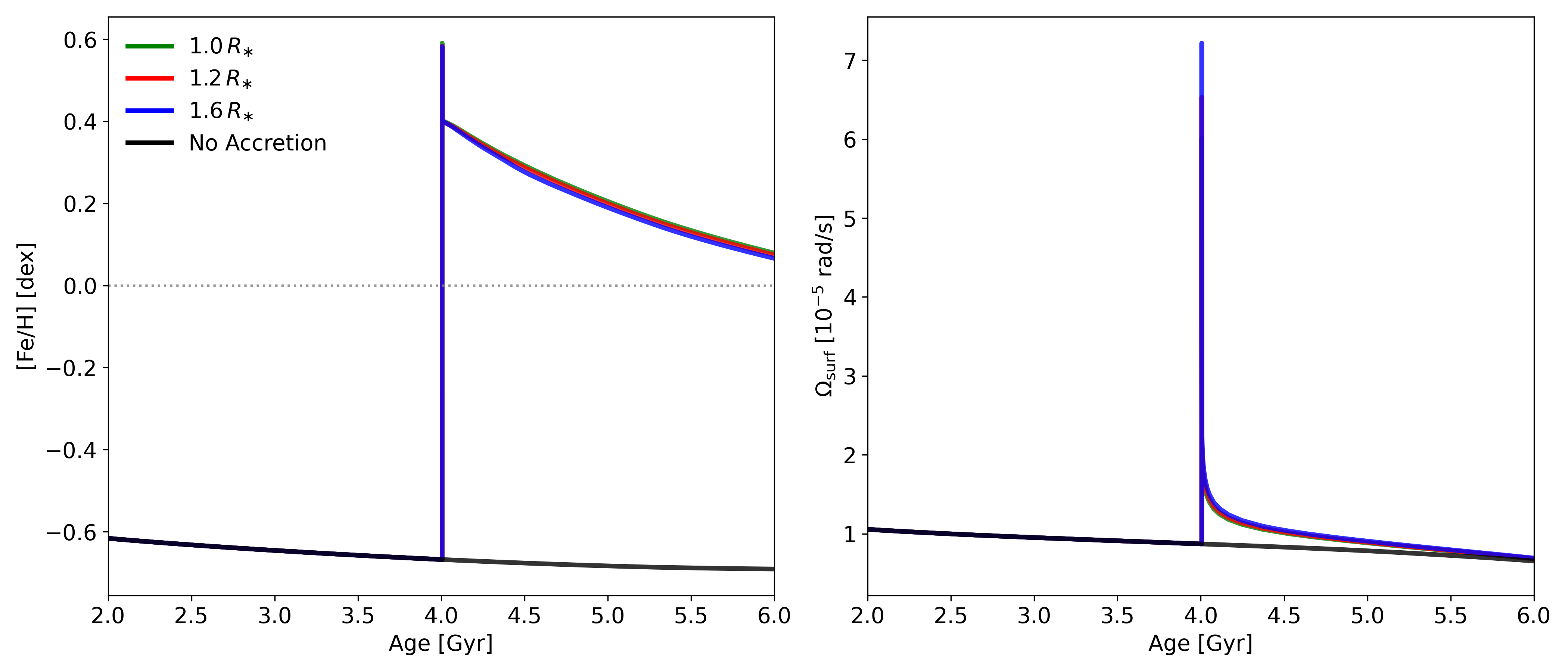}
              \caption{The temporal evolution of the host star's [Fe/H] (left panel) and surface angular velocity (right panel) for different planet engulfment scenarios. The black lines denote the standard stellar evolution models without any engulfment events. The colored lines show the impact of swallowing a rocky planet of 10 $M_\oplus$ at an age of 4.0 Gyr, with variations in the disruption distance: 1.0 $R_*$ (green), 1.2 $R_*$ (red), and 1.6 $R_*$ (blue). Star mass is 1.0 $M_\odot$ and initial metallicity is 0.005.}
              \label{fig:position_test}
         \end{figure*}
         \begin{figure*}
              \sidecaption
              \includegraphics[width=12cm]{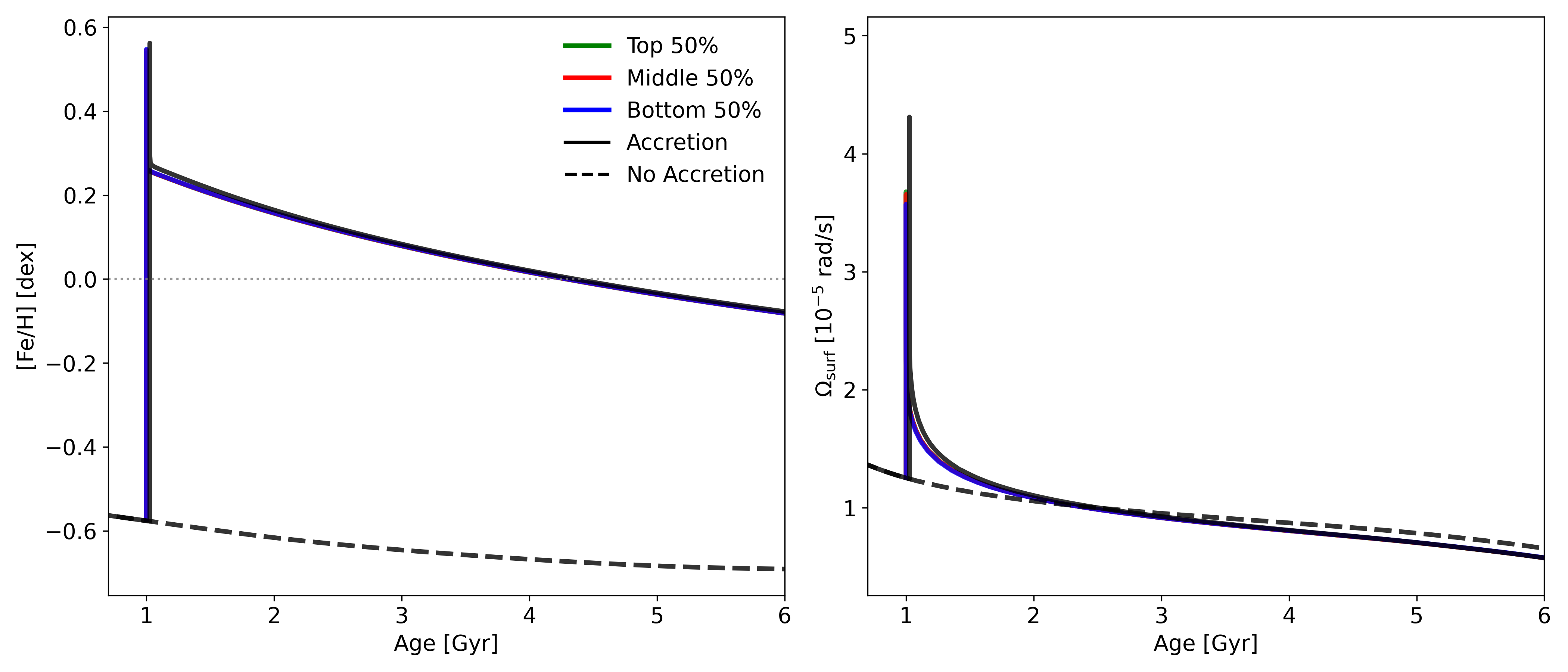}
              \caption{Evolution of [Fe/H] (left) and angular velocity (right) following the injection of $10~M_{\oplus}$ of material into the star's convective envelope. The different colored lines correspond to different injection depths within the convective envelope: green represents the upper half of the grid cells, red represents the middle half, and blue represents the lower half. The black solid and dashed lines represent the accretion and non-accretion models, respectively. The stellar mass is 1.0 $M_{\odot}$ and metallicity is 0.005}
              \label{fig:injection}
        \end{figure*}
            Interestingly, this discrepancy vanishes in the $50\,M_\oplus$ case. We attribute this to the fact that while rotation typically promotes mixing, the associated strong horizontal turbulence can substantially attenuate the intensity of thermohaline mixing and overshooting \citep{2012ApJ...753...49V}. This competition effectively homogenizes the outcomes between rotating and non-rotating models at high accretion masses.
            However, this anomaly is absent in rotating models, which emphasizes the importance of incorporating rotation to accurately depict the chemical and rotational evolution after planet engulfment.

    \subsection{Sensitivity of metallicity and angular momentum change to the accretion radius}\label{subsec:locations}
       For the models in Section \ref{sec:Results}, we adopt a disruption radius of $1.44\,R_*$. However, the theoretical Roche limit varies for planets with different density (see Equation (\ref{eq:roche_limit})). Here, we conducted a sensitivity analysis to assess the impact. Figure \ref{fig:position_test} illustrates scenarios where planetary disruption around $1.0 M_{\odot}$ with 0.005 metallicity occurs at $1.0\,R_*$, $1.2\,R_*$ and $1.6\,R_*$. The results demonstrate that the precise location of disruption has a negligible effect on both the stellar metallicity enrichment and the rotational evolution. This confirms that the variations in angular momentum input caused by the choice of disruption radius are minor, justifying the use of our fixed fiducial value in the broader analysis. Considering that lower-mass and higher-metallicity stars possess thicker convective envelopes, the impact of orbital angular momentum at different locations would only be smaller for them.
        
           Additionally, the planets may disrupt inside the star's convective envelope as presented by \cite{2026ApJ..1003...67L}. While a more robust theory is required to determine whether planetary disruption occurs inside or outside the star, we here examine the impact of internal planetary disruption on the stellar evolution. We tested depositing $10~M_{\oplus}$ of planetary material into the upper, middle, and lower regions of the convective envelope and compare the outcomes with those of the accretion case. Because the injected mass is far smaller than the mass of the convective region  (see Figure \ref{fig:discussion2}), we simply adjusted the elemental abundances at the corresponding grid locations. The angular momentum injection was modeled by assigning only the Keplerian angular momentum of the planetary material to the grid cells where it was added. As shown in Figure \ref{fig:injection}, whether considering continuous accretion or disruption in the convective zone, the long-term evolution of both stellar metallicity enrichment and stellar rotation remains virtually identical, thereby validating the reasonableness of adopting the accretion scenario.

    \subsection{Limitations}\label{subsec:limitations}
          While this study establishes a foundational framework for understanding the chemical and rotational response of stars to planetary engulfment. We have only focused on the moment the star begins to accrete the planetary material, without yet considering the detailed orbital evolution of the planet or the planet-star interactions. In reality, the orbital decay of a planet—driven by tidal dissipation determines the rate and duration of material ingestion. Incorporating these dynamical processes into our work remains a crucial objective for our future simulations.

          This paper primarily focuses on how the accreted material is mixed within the stellar interier, as well as how the resulting chemical enrichment and spin-up signatures gradually diminish as the star envolves. We thus model planetary engulfment as a gradual accretion onto the stellar surface because that was how MESA could handle mass gain. However, denser planets (such as iron-rich bodies) or planets on highly eccentric orbits may plunge intact into the stellar interior, potentially disrupting completely only after penetrating into the convective zone. Accurately modeling this process requires more comprehensive planetary engulfment frameworks.

            This study focuses exclusively on rocky bodies and neglects light elements like Lithium. Their inclusion would offer a crucial "age-sensitive" signature, as the same mixing that dilutes metals also accelerates the destruction of accreted Li in hotter interior layers. However, gas giants are prime candidates for engulfment due to their proximity—possess massive gaseous envelopes subject to hydrodynamic escape and tidal stripping \citep{2010ApJ...712.1107G}. While our characterization of refractory material provides a necessary baseline for modeling the metal-rich cores of giant planets, future models must incorporate the simultaneous accretion of hydrogen/helium-rich gas, which may counteract the mean molecular weight ($\mu$) increase and modify the onset of thermohaline mixing.

\section{Conclusion}\label{sec:Conclusion}

           Planetary engulfment is the ultimate phase of star-planet interaction. Studying this process provides critical insights into how these interactions influence stellar evolution. Using the \textbf{MESA} code, we have systematically investigated the chemical and dynamical response of stars to planetary engulfment. Our simulations reveal that the long-term visibility of accretion signatures is a complex function of the host star's internal structure and the timing of the event.

           We find that the long-term visibility of accretion signatures is defined by a delicate balance between signal intensity and duration. While enrichment intensity is dictated by the thickness of the convective envelope, its duration is mainly governed by the star's MS lifetime. Furthermore, the introduction of stellar rotation increases the complexity of this relationship, as rotational mixing can significantly accelerate the inward transport of accreted material, particularly in stars with thinner envelopes. These effects create a critical trade-off characterized by the long-lived but diluted signals in low-mass stars ($0.7~M_{\odot}$) versus the intense but short-lived enrichment in more massive stars ($1.2~M_{\odot}$) due to rotational effects and short MS lifetimes. Consequently, late-F or early-G type $1.0~M_{\odot}$ stars emerge as the optimal targets, providing the strategic compromise between robust signal intensity and a substantial MS lifetime.

           In addition, the timing of the engulfment event also plays an important role in defining the observable enrichment signal. While ZAMS accretion is rapidly diluted, post-ZAMS events produce more pronounced surface anomalies. For instance, while a $1.0\,M_{\odot}$ star ($Z = 0.005$) ingesting $10\,M_{\oplus}$ at the ZAMS experiences a rapid decline in surface metallicity to sub-solar levels, the same event occurring at an age of 1 Gyr allows the star to maintain a metallicity above solar level as a metal-rich imposter for the subsequent 2 Gyr.

           Finally, non-rotating models predict anomalous chemical abundance variations when accreting planets of different masses. However, rotating models do not exhibit the same behavior, which highlights the importance of using rotating models to characterize the process of planet engulfment.

\begin{acknowledgements} 
           The authors would like to thank the referee for the constructive comments and valuable suggestions, which have significantly improved the quality of this manuscript.
           The authors gratefully acknowledge the ``PHOENIX Supercomputing Platform" jointly operated by the Binary Population Synthesis Group and the Stellar Astrophysics Group at Yunnan Observatories, Chinese Academy of Sciences. The authors would like to express their gratitude to Jason Sevilla for making the open-source code for \textit{Long-Term Lithium Abundance Signatures Following Planetary Engulfment} publicly available, which greatly facilitated the development of this work.

           This work is supported by the National Natural ScienceFoundation of China(grant Nos. 12433009, 12288102 and 12503044). We also acknowledge support from International Centre of Supernovae (ICESUN), Yunnan Key Laboratory of Supernova Research (No. 202505AV340004).

           J.H.G is thankful for the support by YUNNAN ADMINISTRATION OF FOREIGN EXPERTS AFFAIRS,grant No. 202505AO120026.
\end{acknowledgements}
\FloatBarrier

\bibliographystyle{aa}
\bibliography{sample701}
\clearpage
\begin{appendix}
    \section{} 

    \begin{figure}[h]
        \includegraphics[width=0.5\textwidth]{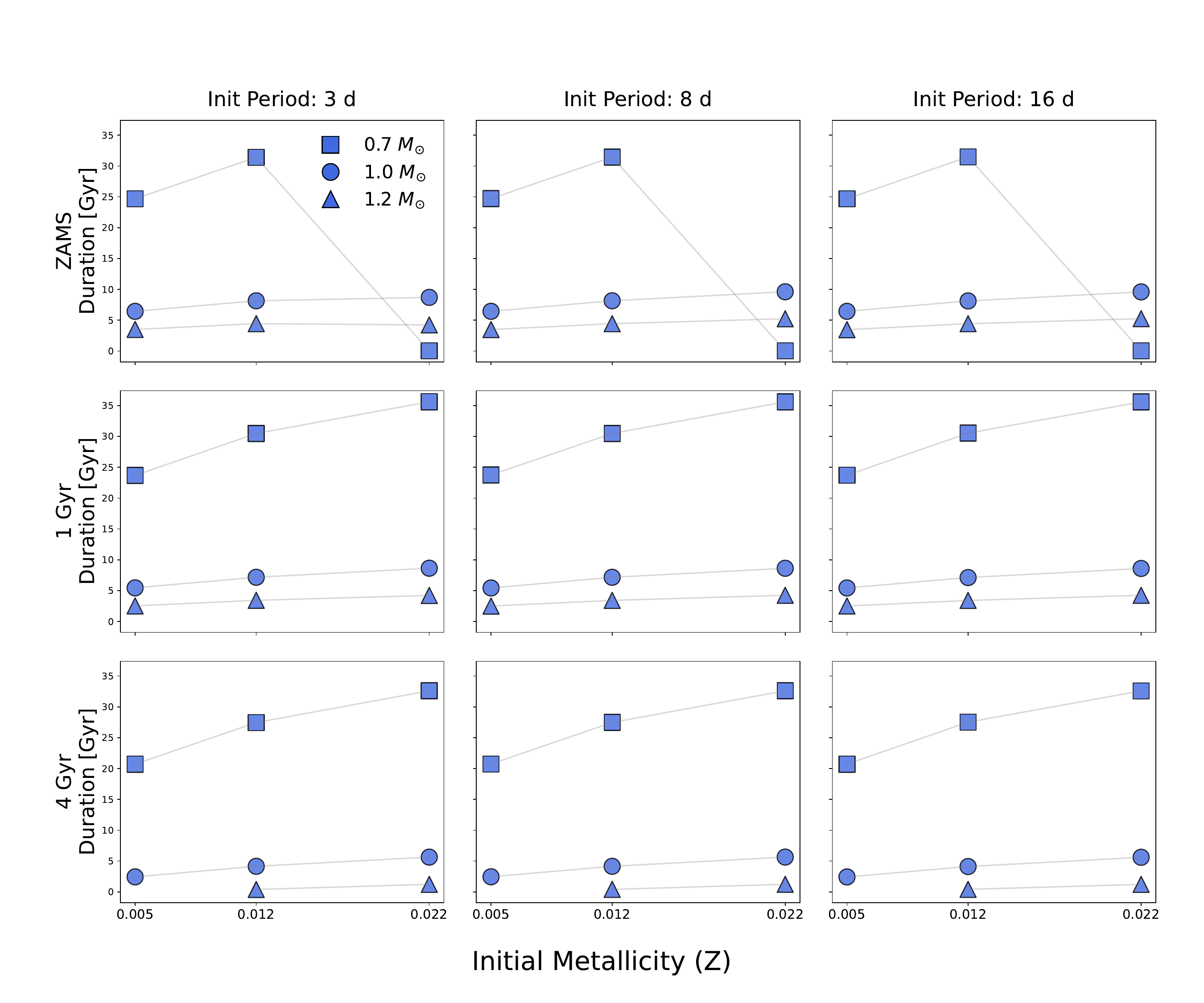}
        \caption{Same as Figure \ref{fig:duration1}, but the engulfed mass of all models is 50 $M_{\oplus}$.}
        \label{fig:duration2}
    \end{figure}
    
    \begin{figure}[h]
        \includegraphics[width=0.5\textwidth]{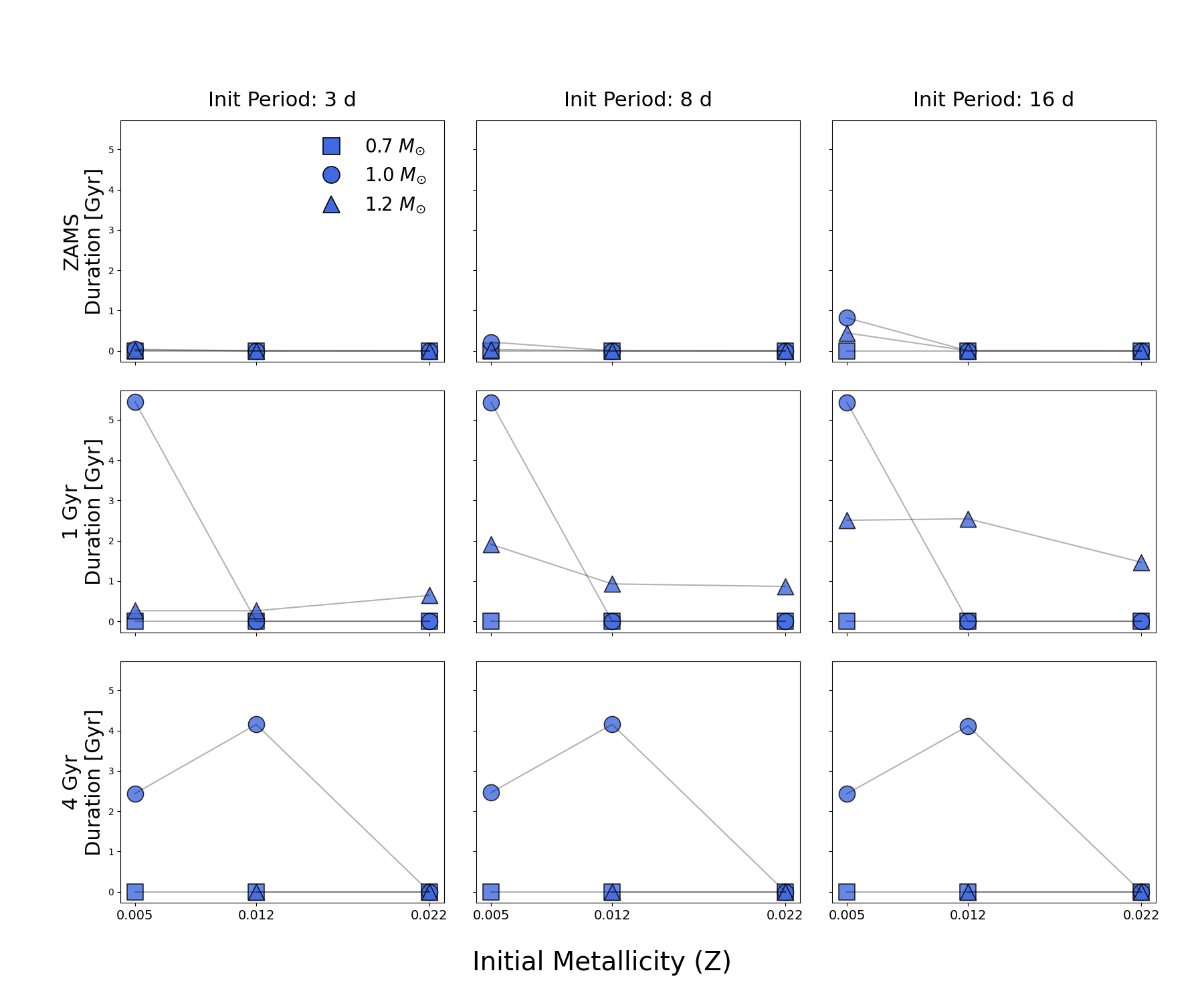}
        \caption{Same as Figure \ref{fig:duration1}, but the threshold for metallicity enrichment was set at 0.3 dex (the accreted masses for the 1.2 $M_\odot$ models with metallicities of 0.012 and 0.022 undergoing engulfment at 4 Gyr are shown in Figure \ref{fig:disruted_mass}).}
        \label{fig:duration3}
    \end{figure}
    
\end{appendix}
\end{document}